\documentclass[12pt,hidelinks]{article}
\pdfoutput=1
\usepackage{mathrsfs,amsmath,epsfig,amssymb,color,graphicx,bm, listings}
\usepackage{algorithm,algorithmic,natbib,subcaption,hyperref,lineno}
\usepackage[margin=1in]{geometry}
\usepackage[symbol]{footmisc}
\usepackage{xr-hyper}
\usepackage{url}
\graphicspath{{figures/}}
\allowdisplaybreaks

\renewcommand{\theequation}{\arabic{section}.\arabic{equation}}
\newtheorem{theorem}{Theorem}
\newtheorem{remark}{Remark}

\newcommand{\bbeta}{{\boldsymbol \beta}}

\newcommand{\bfeta}{{\boldsymbol \eta}}

\newcommand{\balpha}{{\boldsymbol \alpha}}
\newcommand{\btheta}{{\boldsymbol \theta}}

\newcommand{\red}{\color{black}}

\begin{document}

\centerline {\Large  \bf  Efficient Adjusted Joint Significance
Test and  Sobel-Type  }
\centerline {\Large\bf   Confidence Interval for Mediation Effect }
\vspace*{0.2in}

\centerline{ { Haixiang Zhang}}
 \vspace*{0.1in}

\centerline{\it \small $^{}$Center for Applied Mathematics, Tianjin University, Tianjin 300072, China}

\footnotetext[1]{Email: haixiang.zhang@tju.edu.cn (Haixiang Zhang).}
\vspace{1cm}

\begin{abstract}
{ Mediation analysis is an important  statistical tool in many research fields, where the joint significance test is widely utilized for examining mediation effects. Nevertheless, the limitation of this mediation testing method stems from its conservative Type I error, which reduces its statistical power and imposes certain constraints on its utility.  The proposed solution to address this gap is the adjusted joint significance test for one mediator, which introduces a novel data-adjusted approach for assessing mediation effects that showcases significant advancements. The method is specifically designed to be user-friendly, thereby eliminating the necessity for intricate procedures.  We further extend the adjusted joint significance test for small-scale mediation hypotheses with family-wise error rate (FWER) control. Additionally, a novel adjusted Sobel-type  confidence  interval is proposed for the mediation effects, demonstrating significant advancements over conventional Sobel's method. The effectiveness of our mediation testing and confidence interval estimation is assessed through extensive simulations, and compared against a multitude of existing approaches. Finally, we present the application of the method to three substantive datasets with continuous, binary and time-to-event outcomes, respectively.

}
\end{abstract}

\section{Introduction}

Mediation analysis plays an  important role in understanding the causal mechanism that an independent variable $X$ affects
a dependent variable $Y$ through an intermediate variable (mediator) $M$. The utilization of mediation analysis is extensively prevalent across various disciplines,  such as  psychology, economics, epidemiology, medicine,  sociology, behavioral science, and many others. From the perspective of methodological development, \cite{BaronandKenny1986} have laid a solid foundation for mediation analysis. Subsequently, numerous studies have been conducted on this subject. Just to name a few: \cite{2004Confidence} constructed the confidence limits for indirect effect by resampling methods.  \cite{Lijuan-SEM-2011}  and \cite{Zhiyong-2013} introduced the estimating and testing methods for mediation effects with censored data and missing data, respectively. \cite{Quantile_MBR-2014} proposed an  inference technique for quantile mediation effects. \cite{Tyler-2017-JRSSB} considered causal mediation analysis with time-varying exposures and mediators.
\cite{Bayesian-SEM-2021} proposed a Bayesian modeling approach for mediation analysis.  \cite{Zhou-JRSSB-2022} introduced a semiparametric estimation method for  mediation analysis with multiple causally ordered mediators. \cite{big-med-2023} used the subsampled double bootstrap and divide-and-conquer algorithms to conduct statistical mediation analysis on large-scale datasets.
\cite{AB-JRSSB-2023} developed an adaptive bootstrap framework that can be applied to the joint significance test of mediation effect. For more results about mediation analysis, we refer to the reviewing papers by \cite{David-2007} and \cite{Preacher-2015}.

The joint significance test is a crucial statistical approach in the field of mediation analysis, which plays a pivotal role in investigating the causal mechanism underlying mediation effects \cite[]{2008Introduction}. However, the main shortcoming of this method is due to the conservative type I error of mediation testing \cite[]{David-PM-2002}, which largely prevents its popularity for practical users. The statistical analysis of the joint significance tests for large-scale mediation hypotheses in genome-wide epigenetic research has been extensively investigated (\citeauthor{Comp-Huang}, \citeyear{Comp-Huang}; \citeauthor{Dai2022-HDMT-JASA}, \citeyear{Dai2022-HDMT-JASA}; \citeauthor{Liu-JASA-DACT-2022}, \citeyear{Liu-JASA-DACT-2022}), whereas these methods are not applicable to single or small-scale mediation hypotheses.  Furthermore,
the confidence interval for the mediation effect is a crucial aspect in mediation analysis, which is highly valuable in comprehending the  mediation mechanism.  The literature on mediation analysis lacks a substantial number of studies focusing on confidence interval estimation.  The Sobel-type (or normality-based) method and Bootstrap are consistently employed to construct confidence intervals of mediation effects. However, these two methods are inadequate when both the pathway effects along $X \rightarrow M$ and $M \rightarrow Y$ equal zero. To improve the performances of joint significance test and Sobel-type confidence interval, we propose two novel data-adjusted mediation analysis methods with theoretical verification. The main advantages of our proposed method are as follows: First, the adjusted joint significance test and  adjusted sobel-type confidence interval
are two flexible and data-driven methods. The implementation of the two proposed methods is particularly convenient from a practical perspective. Specifically, our method ensures user-friendliness by eliminating complex procedures.  Second, the test method we propose exhibits significant advancements in terms of size and power when compared to the conventional joint significance test.  The enhanced powers are particularly evident for those mediation effects that are relatively weak, making them challenging to be recognized as significant mediators by traditional methods.  Third, the explicit formulation of the coverage probability and length of the adjusted Sobel-type confidence interval is provided for comparison with conventional Sobel's method.

The remainder of this paper is organized as follows: In Section \ref{sec-2}, we review some details about the traditional joint significance test for mediation effects. Then we propose a novel data-adjusted
 joint significance test for one mediator, together with the explicit expression of size. Meanwhile, an adjusted Sobel test is also introduced, which shows a significant improvement compared to traditional Sobel's method.
 In Section \ref{sec-3}, we implement the adjusted joint significance test towards small-scale multiple testing with FWER control. Section \ref{section-4-ASobel} introduces an adjusted Sobel-type method to construct confidence interval for the mediation effect. Section \ref{sec-5} presents some simulation studies to assess the performance of our method. In Section \ref{sec-6}, we perform mediation analysis for three real-world datasets with the proposed method. Some concluding
 remarks are provided in Section \ref{sec-7}. All proof details of theorems are presented in the Supplementary Material. Finally, our method offers a publicly available and user-friendly R package, called $\tt AdjMed$, which can be accessed at \url{https://github.com/zhxmath/AdjMed}.

\begin{figure}[htp]
\centerline{\includegraphics[scale=0.35]{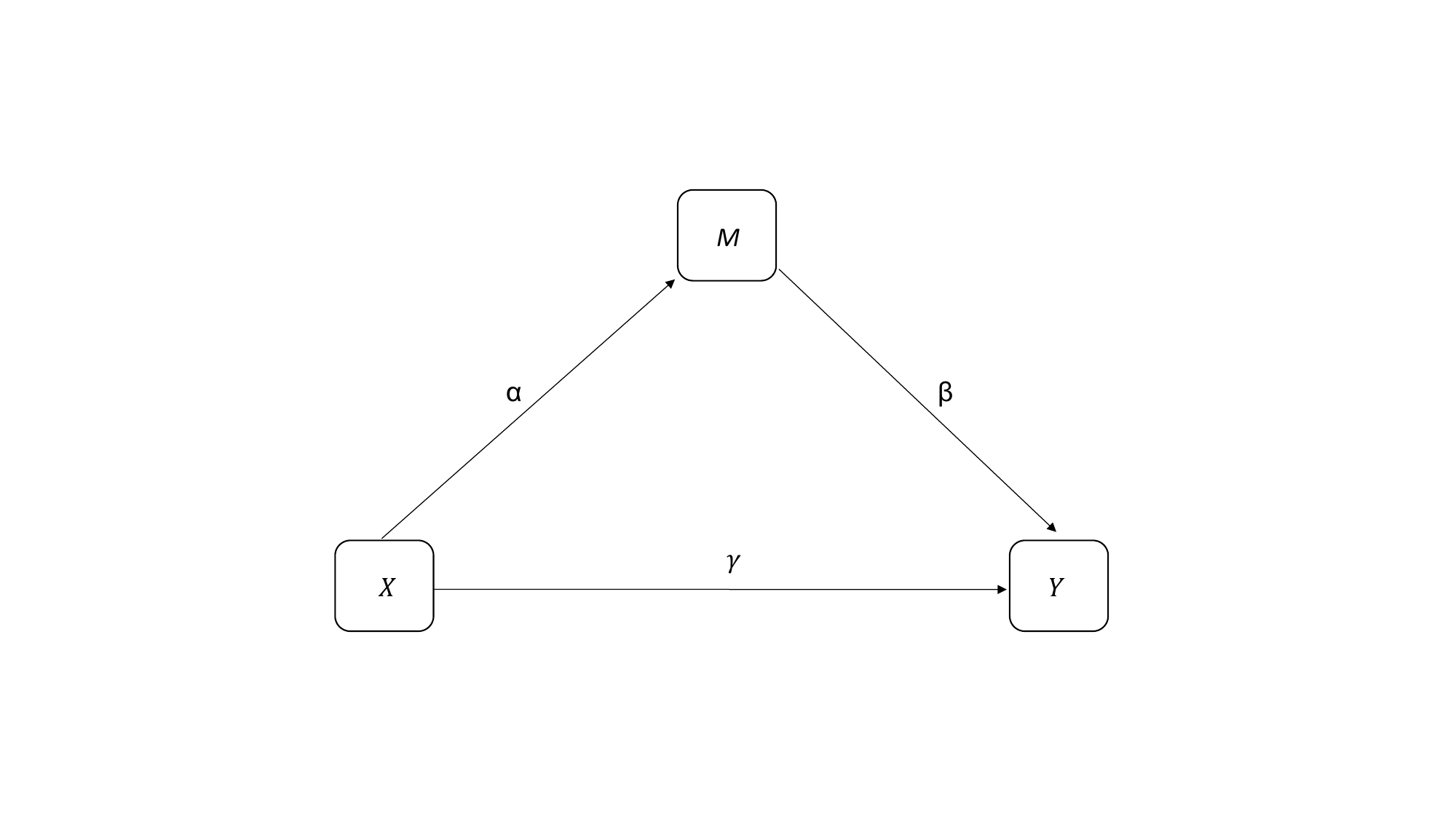}}
 \vspace{-1.5cm}
\begin{center}
\caption{A scenario of  mediation  model with one mediator (confounding variables are omitted).}\label{fig:1}
\end{center}
\end{figure}

\section{Adjusted Joint Significance Test}\label{sec-2}
\setcounter{equation}{0}
To begin with, we review some fundamental notations within the context of mediation analysis. Let $X$ be an exposure,  $M$ be the mediator and $Y$ be the outcome (see Figure \ref{fig:1}). As described by \cite{2008Introduction}, the aim of  mediation analysis is focused on investigating the causal mechanism along the pathway $X\rightarrow M \rightarrow Y$. Generally speaking, the causal effect $X\rightarrow M$ is parameterized by $\alpha$ (after
adjusting for confounders), and the causal effect $M \rightarrow Y$  is parameterized by $\beta$ (after adjusting for exposure and confounders).  The mediating effect of $M$ in this case is described by $\alpha\beta$, commonly  known as the ``product-coefficient" approach \cite[]{David-PM-2002}. To evaluate whether $M$ plays an intermediary role between $X$ and $Y$, it is customary to conduct hypothesis testing at a significance level of $\delta$ as follows:
\begin{eqnarray}\label{Eq-2.1}
H_0:~\alpha\beta=0~\leftrightarrow~ H_A: \alpha\beta\neq 0.
\end{eqnarray}
The rejection of the null hypothesis would indicate that $M$ is a statistically significant mediator in the pathway $X\rightarrow M \rightarrow Y$. The null hypothesis $H_0$ is worth noting as it is composite. Specifically, $H_0$ can be decomposed equivalently into the union of three disjoint component null hypotheses $H_0 = H_{00} \cup H_{01} \cup H_{10}$, where
\begin{eqnarray*}
&&H_{00}:~\alpha=0, \beta=0;\\
&&H_{01}:~\alpha=0, \beta\neq0;\\
&&H_{10}:~\alpha\neq0, \beta=0.
\end{eqnarray*}

Let $T_\alpha= {\hat{\alpha}}/{\hat{\sigma}_{\alpha}}$ and $T_\beta = {\hat{\beta}}/{\hat{\sigma}_{\beta}}$ be the statistics for testing $\alpha=0$ and $\beta=0$.
Here $\hat{\alpha}$ and $\hat{\beta}$ are the estimates for $\alpha$ and $\beta$, respectively; $\hat{\sigma}_{\alpha}$ and $\hat{\sigma}_{\beta}$ are the
estimated standard errors of $\hat{\alpha}$ and $\hat{\beta}$, respectively.
Under the null hypothesis, as the sample size $n$ tends to infinity, it can be observed that
\begin{eqnarray}\label{Eq2-3}
T_\alpha \stackrel{\mathcal{D}}{\longrightarrow}  N(0,1)~{\rm and}~T_\beta \stackrel{\mathcal{D}}{\longrightarrow} N(0,1),
\end{eqnarray}
where $\stackrel{\mathcal{D}}{\longrightarrow}$ denotes convergence in distribution. The corresponding p-values for $T_\alpha$ and $T_\beta$ are
\begin{eqnarray}
P_\alpha &= &2\{1-\Phi_{N(0,1)}(|T_\alpha|)\},\label{Eq2-4}\\
P_\beta &=& 2\{1-\Phi_{N(0,1)}(|T_\beta|)\},\label{Eq2-5}
\end{eqnarray}
where $T_\alpha$ and $T_\beta$ are defined in (\ref{Eq2-3}), $\Phi_{N(0,1)}(\cdot)$ is the cumulative distribution function of $N(0,1)$. The joint significance (JS) test, also known as the MaxP test, is widely recognized as one of the most popular methods for mediation analysis in the field.  The purpose of the JS test is to reject $H_0$ when both $\alpha = 0$ and $\beta = 0$ are simultaneously rejected. The JS test statistic is 
\begin{eqnarray}\label{Eq3-1}
P_{JS} = \max(P_\alpha, P_\beta),
\end{eqnarray}
where $P_{\alpha}$ and $P_\beta$ are given in (\ref{Eq2-4}) and (\ref{Eq2-5}), respectively. The practical applicability of JS test has led to its widespread adoption across various research fields, including the social and biomedical sciences.  However, the JS test suffers from overly conservative type I error,  especially when  both $\alpha =0$ and $\beta =0$ \cite[]{David-PM-2002}. In the literature,  this is a long-existed and unresolved problem for the JS test. To bridge this gap, we aim to propose a novel data-adjusted method for improving the statistical efficiency of the JS test, especially focusing on the conservative issue in the case of $\alpha=\beta=0$.

 Under $H_0$, the traditional JS test regards $P_{JS}$ as a uniform random variable over $(0,1)$. i.e., $P_{JS} \sim U(0,1)$. However, the actual distribution of $P_{JS}$ is not $U(0,1)$ under the component null hypotheses $H_{00}$, which is the reason of  the conservative performance of traditional JS test. To provide further insights on this matter, let us consider $Z_1$ and $Z_2$ as two independent random variables that follow a uniform distribution $U(0,1)$. Let ${Z} = \max ({Z_1}, {Z_2})$ and its density function is  $f_{{Z}}(z)$. We can derive the distribution function of $Z$ as
\begin{align*}
F_Z(z) = \mathbb{P}(Z \leq z) = \mathbb{P}(Z_1 \leq z, Z_2\leq z)= \mathbb{P}(Z_1 \leq z)\mathbb{P}(Z_2\leq z)=z^2,
\end{align*}
 i.e., the density function of $Z$ is $f_Z(z) = \frac{1}{2}z$ for $0\leq z \leq 1$ and $f_Z(z) = 0$, otherwise. Therefore, the maximum of two  independent p-values ($P_{JS}$) does not follow $U(0,1)$. This provides a theoretical view about the conservative performance of JS test when
both $\alpha = 0$ and $\beta =0$. Note that the distribution function of $Z^2 = \{\max ({Z_1}, {Z_2})\}^2$ is
\begin{eqnarray}\label{EQ-max}
F_{Z^2}(z) &=& \mathbb{P}(Z^2 \leq z)\nonumber\\
&=& \mathbb{P}(Z_1 \leq z^{1/2})\mathbb{P}(Z_2 \leq z^{1/2})\nonumber\\
&=&\left\{\begin{array}{ll} 0,&z<0,\\
z,&0\leq z < 1,\\
1,&z\geq 1.
\end{array}\right.
\end{eqnarray}
Under $H_{00}$, the squared maximum of two p-values ($P^2_{JS}$) has an uniform distribution over (0, 1). This discovery provides insight into resolving the conservative challenge of conventional JS test.

\begin{figure}[htp] 
  \centering
  \begin{subfigure}{0.45\textwidth}
    \includegraphics[width=\textwidth]{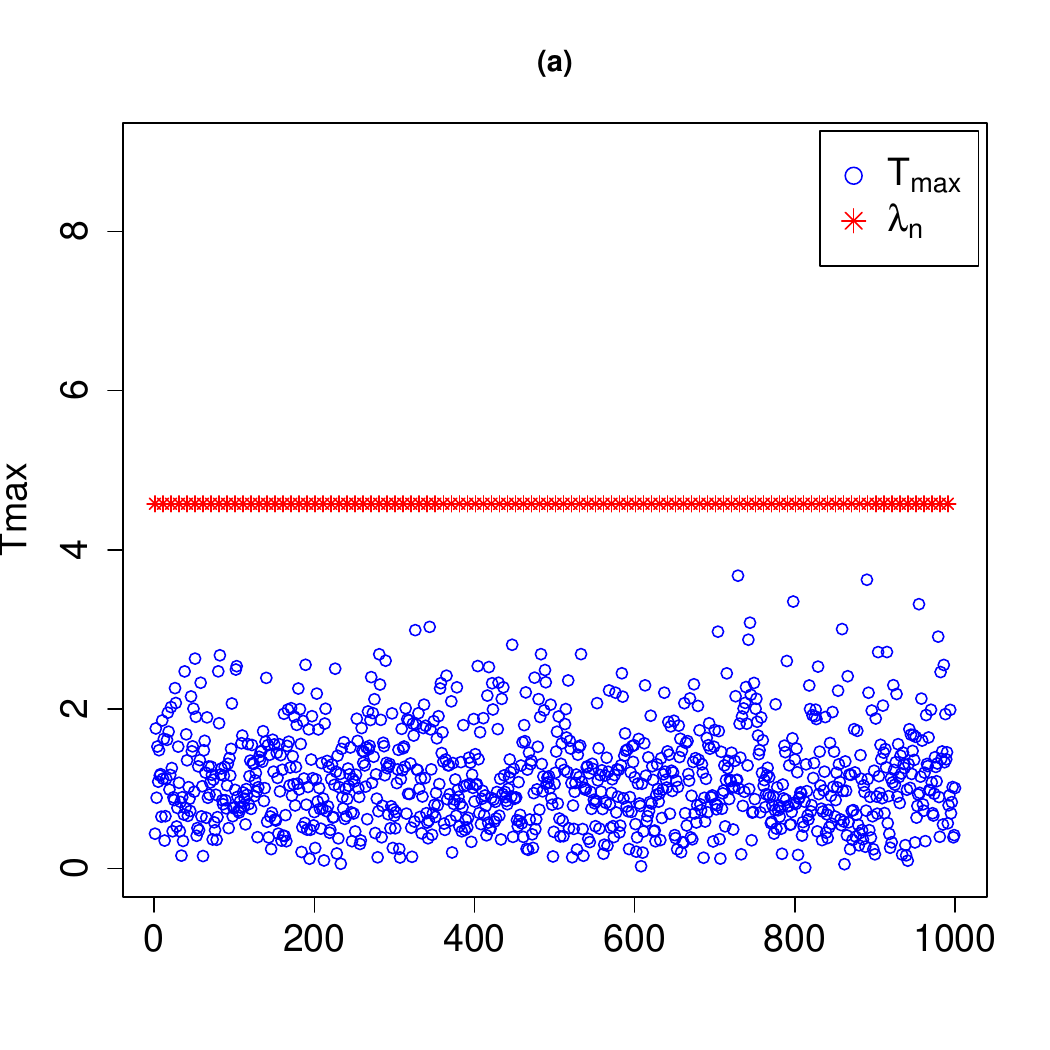}
    \caption{The parameter is $(\alpha, \beta) = (0,0)$.}
  \end{subfigure}
  \begin{subfigure}{0.45\textwidth}
    \includegraphics[width=\textwidth]{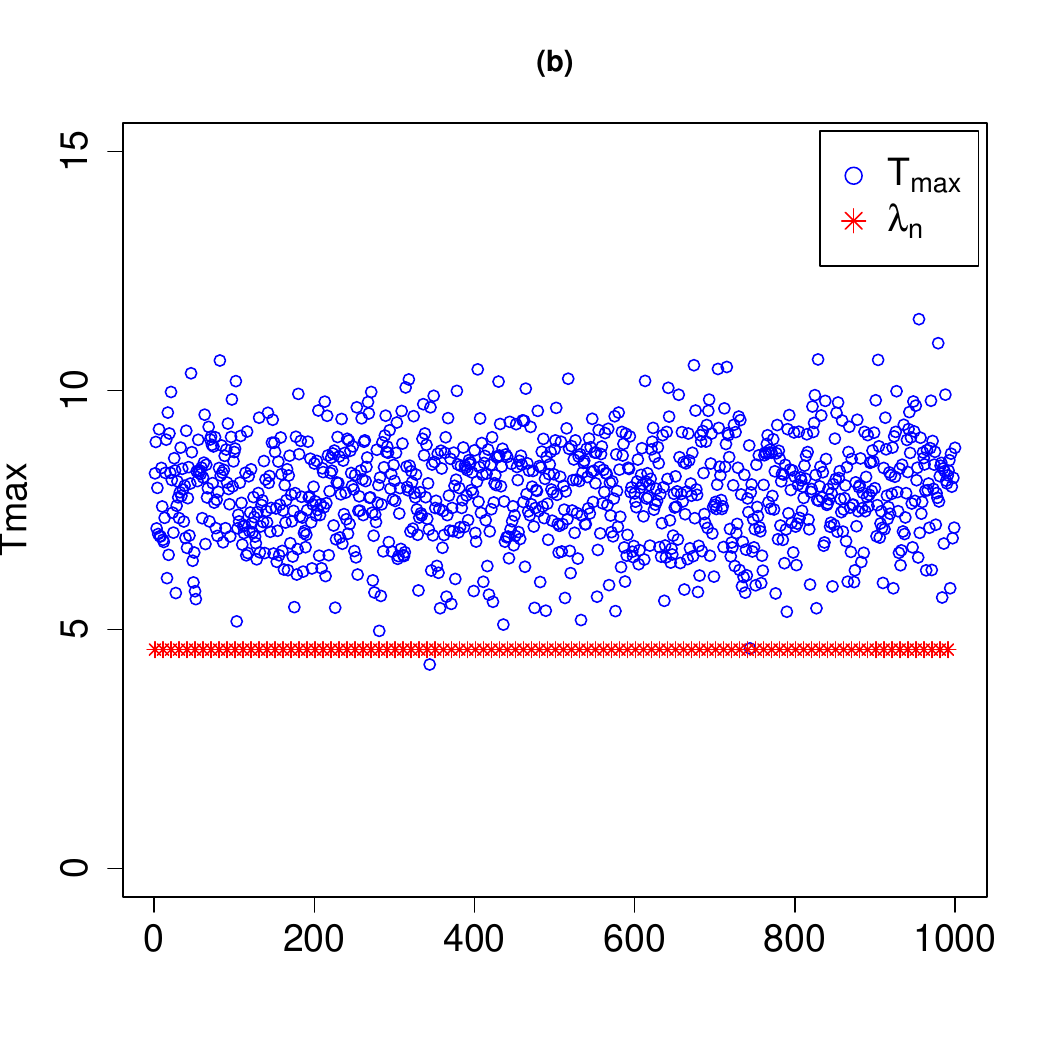}
    \caption{The parameter is $(\alpha, \beta) = (0.25,0)$.}
  \end{subfigure}
 \vspace{-0.1cm}
\begin{center}
  \caption{The scatter plots of $T_{max}=\max\{|T_\alpha|, |T_\beta|\}$ and $\lambda_n=\sqrt{n}/\log(n)$ with 1000 repetitions.}
  \label{fig:2}
\end{center}
\end{figure}

To address the conservatism of JS test, we attempt to differentiate $H_{00}$ from $H_{01}$ and $H_{10} $ reasonably using a specific criterion. Under $H_{00}$, we have $\lim_{n\rightarrow \infty}\mathbb{P}(|T_\alpha|< \lambda_n, |T_\beta|< \lambda_n|H_{00})=1$, where $\lambda_n = o(\sqrt{n})$ and $\lambda_n\rightarrow \infty$ as the sample size $n\rightarrow \infty$. However, $\lim_{n\rightarrow \infty}\mathbb{P}(|T_\alpha|< \lambda_n, |T_\beta|< \lambda_n|H_{10})=0$ and  $\lim_{n\rightarrow \infty}\mathbb{P}(|T_\alpha|< \lambda_n, |T_\beta|< \lambda_n|H_{01})=0$. The null hypothesis $H_0$ is more likely to be regarded as $H_{00}$ instead of $H_{01}$ or $H_{10}$  when  $\max\{|T_\alpha|, |T_\beta|\} < \lambda_n$.  The approach employed by \cite{AB-JRSSB-2023} in constructing bootstrap-based testing statistics for evaluating mediation effects should be noted, while our focus lies on the development of innovative testing statistics that eliminate the need for re-sampling procedures. The above-mentioned concept is exemplified through a numerical illustration provided by us.   Specifically,   we generate a series of random samples from the linear mediation models: $Y= 0.5 + 0.5X + \beta M + \epsilon$ and $M= 0.5 + \alpha X + e$, where $X$, $\epsilon$ and $e$ follow from  $N(0,1)$. The resulting $T_\alpha$ and $T_\beta$ are obtained by the ordinary least square method. In Figure \ref{fig:2}, we present the scatter plots of $T_{max}=\max\{|T_\alpha|, |T_\beta|\}$ and $\lambda_n=\sqrt{n}/\log(n)$ with 1000 repetitions,  where  the sample size is $n=1000$, the parameters $(\alpha, \beta)$ are chosen as $(\alpha, \beta) = (0,0)$, and (0.25, 0),  respectively. The observation from Figure \ref{fig:2} suggests that, under the assumption of $\alpha=\beta=0$, there is a high probability that $T_{max}$ is significantly smaller than $\lambda_n$ as $n\rightarrow \infty$. However, $T_{max}$ is asymptotically much larger than $\lambda_n$ when $\alpha\neq0$, and the same conclusion also applies to $\beta\neq0$.

\begin{figure}[htp] 
  \centering
  \begin{subfigure}{0.4\textwidth}
    \includegraphics[width=\textwidth]{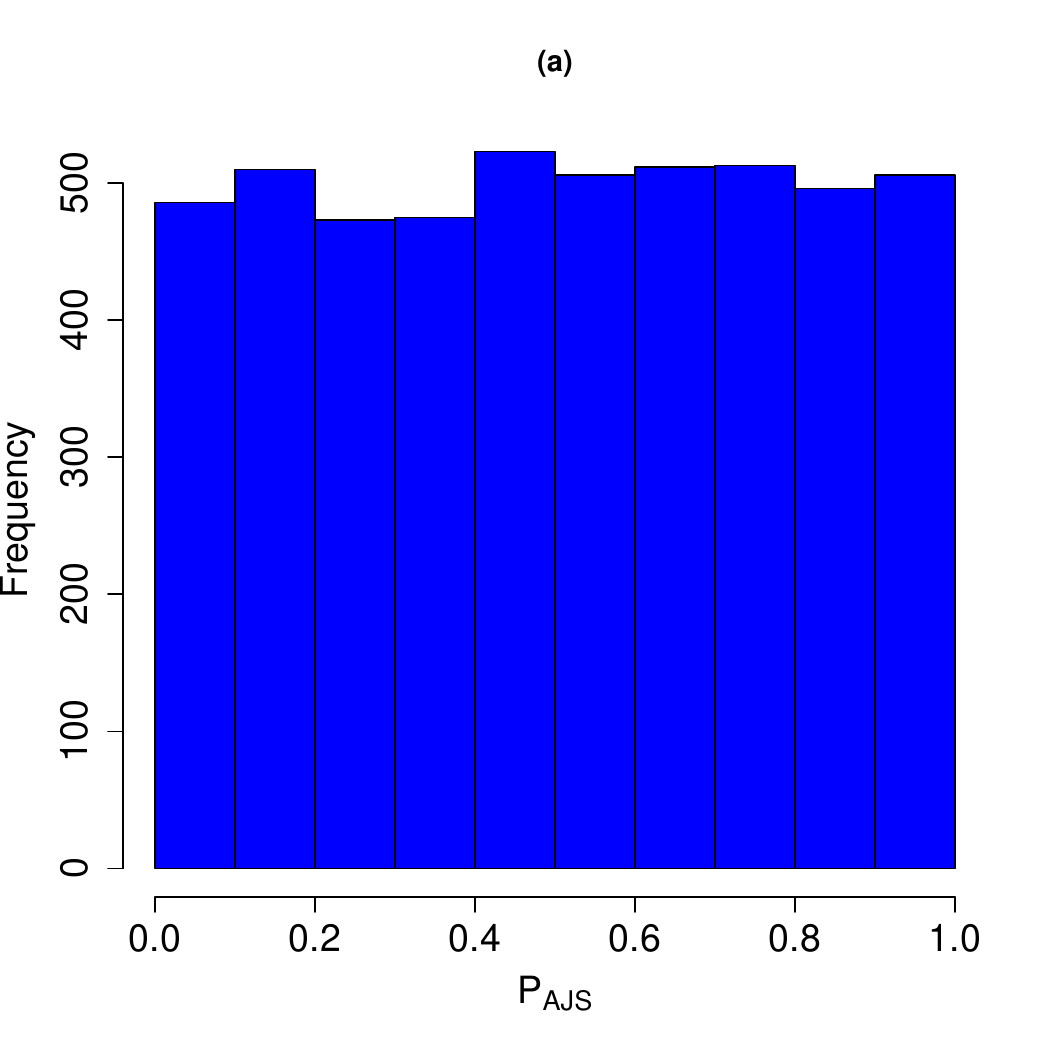}
    \caption{The p-values of AJS method.}
  \end{subfigure}
  \begin{subfigure}{0.4\textwidth}
    \includegraphics[width=\textwidth]{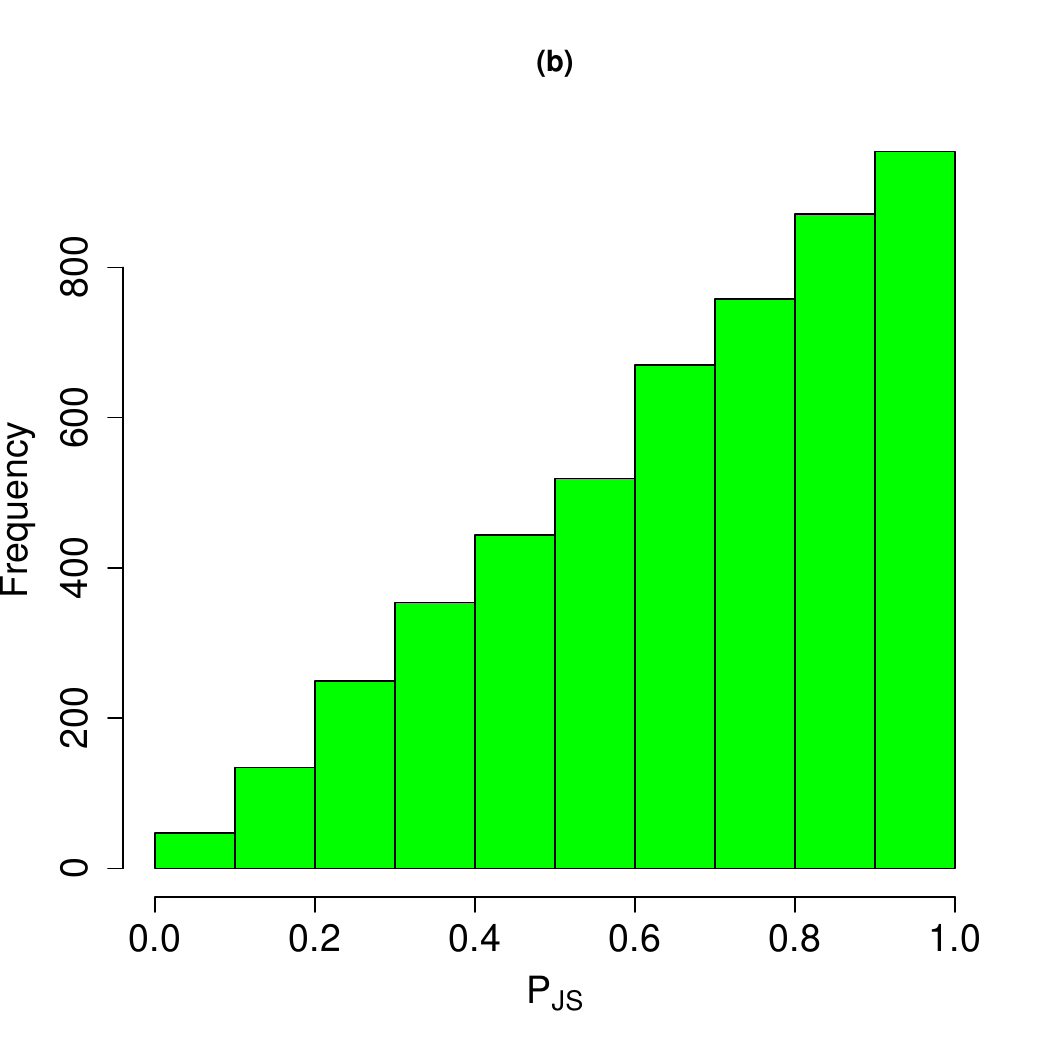}
    \caption{The p-values of JS method.}
  \end{subfigure}
 \vspace{-0.1cm}
\begin{center}
  \caption{The histogram of p-values for $(\alpha, \beta) = (0, 0)$ with $\lambda_n =  \sqrt{n}/\log(n)$.}
  \label{fig:3}
\end{center}
\end{figure}

Motivated by the aforementioned findings, we propose a novel adjusted joint significance (AJS)  test procedure for (\ref{Eq-2.1}), where the p-value is defined as
\begin{eqnarray}\label{Eq3-2}
P_{AJS}
&=&\left\{\begin{array}{ll} P_{JS},& \max\{|T_\alpha|, |T_\beta|\} \geq \lambda_n,\\
P^2_{JS},&\max\{|T_\alpha|, |T_\beta|\} < \lambda_n.
\end{array}\right.
\end{eqnarray}
Here, the threshold  is chosen as $\lambda_n = \sqrt{n}/\log(n)$ satisfying $\lambda_n = o(\sqrt{n})$ and $\lambda_n\rightarrow \infty$ as $n\rightarrow \infty$; $T_{\alpha}$, $T_{\beta}$ and $P_{JS}$  are given in (\ref{Eq2-3}) and (\ref{Eq3-1}), respectively. 
 In Figure \ref{fig:3}, we provide an illustrative example of the p-values for AJS and JS methods with $(\alpha, \beta) = (0, 0)$ and $n=2000$. The data are generated from the same models of Figure \ref{fig:2} with 5000 repetitions. The distribution of $P_{AJS}$ is observed to be uniform in Figure \ref{fig:3}, whereas the histogram indicates a right skewness in the distribution of $P_{JS}$. Meanwhile, the Figure \ref{fig:3} provides an intuitive explanation for the conservatism of traditional JS method when $\alpha=\beta=0$.

The decision rule of AJS test is that the null hypothesis $H_0$ is rejected if $P_{AJS}$ is much smaller than the significance level $\delta$, where $P_{AJS}$ is defined in (\ref{Eq3-2}).
The explicit expression of size in Theorem \ref{Th1} is derived, providing a valuable tool for ensuring the rationality of the AJS test.

\begin{theorem}\label{Th1}
Given that $\lambda_n = o(\sqrt{n})$ and as the sample size $n$ approaches infinity, {with probability approaching one}, under $H_{00}$ the asymptotic size of our proposed AJS test  is
\begin{eqnarray}\label{SJ00}
\lim\limits_{n\rightarrow \infty}{ Size(AJS|H_{00})} =  \delta,
\end{eqnarray}
where $\delta$ is the significance level.  Under $H_{10}$,  {with probability approaching one}, the asymptotic size of the AJS test satisfies
\begin{eqnarray}\label{SJ10}
\lim\limits_{n\rightarrow \infty}{ Size(AJS|H_{10})} &=& \delta.
\end{eqnarray}
 Similarly,  {with probability approaching one}, under $H_{01}$ the asymptotic size of the AJS test is $$\lim_{n\rightarrow \infty}{Size(AJS|H_{01})} = \delta.$$ 
\end{theorem}

Theoretically, we conduct a comparative analysis between AJS and traditional JS in terms of size.
Under $H_{00}$, the size of traditional  JS test for (\ref{Eq-2.1}) is
\begin{eqnarray*}
Size(JS|H_{00}) &=& \mathbb{P}(P_{JS} < \delta | H_{00})\\
&=& \mathbb{P}(P_{\alpha} < \delta | H_{00})\mathbb{P}(P_{\beta} < \delta | H_{00})\\
&=& \delta^2,
\end{eqnarray*}
which explains the conservative phenomenon of traditional JS test.

The Sobel test \cite[]{Sobel-1982}  is another widely used method for mediation analysis in the field, and the corresponding  test statistic is
\begin{eqnarray}\label{Eq2-2}
T_{Sobel} = \frac{\hat{\alpha}\hat{\beta}}{\{\hat{\alpha}^2\hat{\sigma}^2_{\beta} + \hat{\beta}^2\hat{\sigma}^2_{\alpha}\}^{1/2}},
\end{eqnarray}
where $\hat{\alpha}$ and $\hat{\beta}$ are the estimates for $\alpha$ and $\beta$, respectively; $\hat{\sigma}_{\alpha}$ and $\hat{\sigma}_{\beta}$ are the
estimated standard errors of $\hat{\alpha}$ and $\hat{\beta}$, respectively. The
decision rule of traditional Sobel test  relies on the standard asymptotic normality
of $T_{Sobel}$. We can reject $H_0$ if the  $p$-value, $P_{Sobel}$,
is smaller than a specified significance level $\delta$, where $$P_{Sobel} = 2\{1-\Phi_{N(0,1)}(|T_{Sobel}|)\},$$ 
$\Phi_{N(0,1)}(\cdot)$ is the cumulative distribution function of $N(0,1)$, and $T_{Sobel}$ is given in (\ref{Eq2-2}).  The Sobel test suffers from overly conservative type I error,  especially when  both $\alpha =0$ and $\beta =0$ \cite[]{David-PM-2002}. From \cite{Liu-JASA-DACT-2022}, the
Sobel statistic $T_{sobel}$ has an asymptotic normal distribution $N(0,1)$ under $H_{01}$ and $H_{10}$, while its asymptotic distribution is $N(0,1/4)$ in the case of $H_{00}$. Therefore, the performance of  traditional Sobel test is conservative when $\alpha =0$ and $\beta =0$.  Similar to the idea of AJS method,  we propose a novel adjusted Sobel (ASobel) test procedure for (\ref{Eq-2.1}), and the p-value is defined as
\begin{eqnarray}\label{Eq2-6}
P_{ASobel}
&=&\left\{\begin{array}{ll} 2\{1-\Phi_{N(0,1)}(|T_{Sobel}|)\},&\max\{|T_\alpha|, |T_\beta|\} \geq \lambda_n,\\
2\{1-\Phi_{N(0,1/4)}(|T_{Sobel}|)\},&\max\{|T_\alpha|, |T_\beta|\} < \lambda_n,
\end{array}\right.
\end{eqnarray}
where $\Phi_{N(0,1/4)}(\cdot)$ is the cumulative distribution function of $N(0,1/4)$; $T_{Sobel}$ is given in (\ref{Eq2-2});
$T_\alpha$, $T_\beta$ and $\lambda_n$ are the same as that of $P_{AJS}$ in (\ref{Eq3-2}). The decision rule of the ASobel test states that the null hypothesis $H_0$ is rejected if the test statistic ($P_{ASobel}$) is significantly smaller than the predetermined significance level $\delta$.  The p-values for ASobel and Sobel methods with $(\alpha, \beta) = (0, 0)$ and $n=2000$ are illustrated in Figure S.1. The sample data are the same as depicted in Figure \ref{fig:3}.  The distribution of $P_{ASobel}$ appears to be uniformly distributed in Figure S.1, while the histogram indicates a right-skewed distribution for $P_{Sobel}$. The Figure S.1 further elucidates the inherent conservatism of the traditional Sobel method when both $\alpha$ and $\beta$ are set to zero.

Given that $\lambda_n = o(\sqrt{n})$ and as  $n\rightarrow \infty$, {with probability approaching one}, under $H_{0}$ the asymptotic size of our proposed ASobel test for (\ref{Eq-2.1}) is
\begin{eqnarray}\label{size-AS-14}
\lim\limits_{n\rightarrow \infty}{ Size(ASobel|H_{0})} = \delta. 
\end{eqnarray}
The proof details of (\ref{size-AS-14}) of ASobel test are presented in the Supplemental Material. The ASobel test, although superior to the conventional Sobel test, demonstrates inferior performance compared to the AJS method in numerical simulations. Consequently, our focus on hypothesis testing does not heavily emphasize the ASobel method. However, the ASobel plays a crucial role in constructing efficient confidence intervals for mediation effects, which will be presented in the Section \ref{section-4-ASobel}.

\begin{figure}[htp]
\centerline{\includegraphics[scale=0.37]{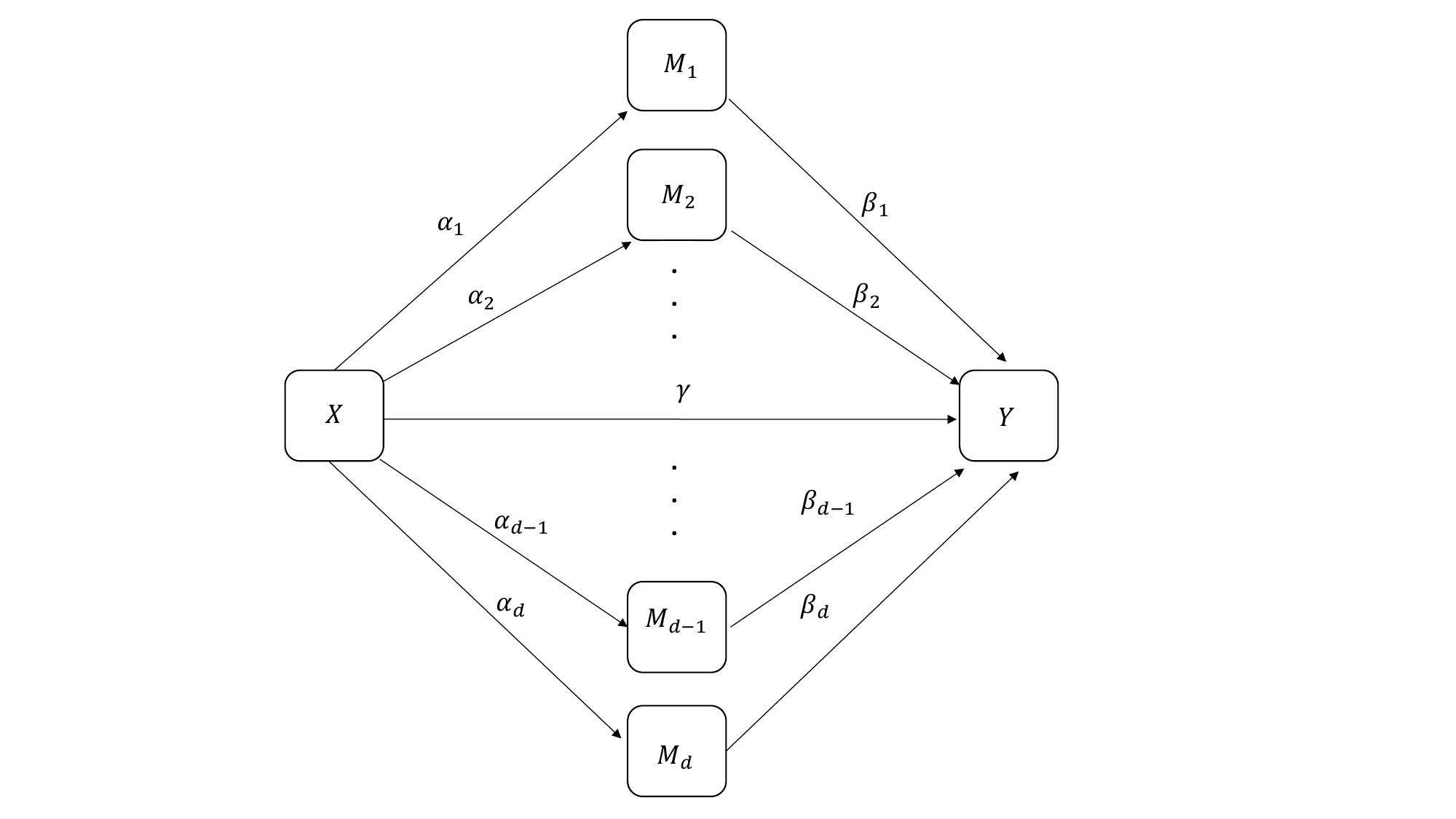}}
 \vspace{-0.2cm}
\begin{center}
\caption{A scenario of  mediation  model with multiple mediators (confounding variables are omitted).}\label{fig:5}
\end{center}
\end{figure}

\section{Multiple Mediation Testing with FWER Control}\label{sec-3}
\setcounter{equation}{0}
The proposed AJS method is extended in this section to address small-scale multiple testing with FWER control, thereby excluding the focus on high-dimensional mediators. 
Let $X$ be an exposure,
$\mathbf{M} = (M_1,\cdots,M_d)^\prime$ be a vector of $d$-dimensional mediators, and $Y$ be an outcome of interest.
Following \cite{2008Introduction}, it is commonly assumed that the causal relations $X \rightarrow M_k$ and $M_k \rightarrow Y$ are parameterized by $\alpha_k$ and $\beta_k$ (see Figure \ref{fig:5}), respectively. Let $\Omega = \{k: \alpha_k\beta_k\neq 0, k=1,\cdots,d\}$ be the index set of significant (or active) mediators. Under the significance level $\delta$,
we are interested in the  small-scale multiple testing problem:
\begin{eqnarray}\label{Eq-4-1}
H^{(k)}_0:~\alpha_k\beta_k=0~~\leftrightarrow~~ H^{(k)}_A: \alpha_k\beta_k\neq 0,~~k=1,\cdots,d,
\end{eqnarray}
where $d$ is not very large (i.e., the $\mathbf{M}$ is not high-dimensional), and each null hypothesis $H^{(k)}_0$ is composite with three components:
\begin{eqnarray*}
&&H^{(k)}_{00}: \alpha_k=0, \beta_k=0;\\
&&H^{(k)}_{10}: \alpha_k\neq 0, \beta_k=0;\\
&&H^{(k)}_{01}: \alpha_k=0, \beta_k\neq 0.
\end{eqnarray*}


First we focus on extending the AJS method (in Section \ref{sec-2}) for the multiple testing in (\ref{Eq-4-1}). For $k=1,\cdots,d$, we denote 
\begin{eqnarray}\label{Eq-4-4}
P_{\alpha_k} = 2\{1-\Phi_{N(0,1)}(|T_{\alpha_k}|)\},~{\rm and}~ P_{\beta_k} = 2\{1-\Phi_{N(0,1)}(|T_{\beta_k}|)\},~k=1,\cdots,d.
\end{eqnarray}
Here $T_{\alpha_k}=\hat{\alpha}_k/\hat{\sigma}_{\alpha_k}$ and $T_{\beta_k}=\hat{\beta}_k/\hat{\sigma}_{\beta_k}$; $\hat{\alpha}_k$ and $\hat{\beta}_k$ are the estimates for $\alpha_k$ and $\beta_k$, respectively. $\hat{\sigma}_{\alpha_k}$ and $\hat{\sigma}_{\beta_k}$ are the
estimated standard errors of $\hat{\alpha}_k$ and $\hat{\beta}_k$, respectively. Specifically, let
\begin{eqnarray}\label{Eq-4-5}
P^{(k)}_{JS} = \max(P_{\alpha_k}, P_{\beta_k}),
\end{eqnarray}
where $P_{\alpha_k}$ and  $P_{\beta_k}$ are given in (\ref{Eq-4-4}), $k=1,\cdots,d$. We propose an AJS test method with the  statistic being defined as
\begin{eqnarray}\label{Eq-4-6}
P^{(k)}_{AJS}
&=&\left\{\begin{array}{ll} P^{(k)}_{JS},& \max\{|T_{\alpha_k}|, |T_{\beta_k}|\} \geq \lambda_n,\\
\{P^{(k)}_{JS}\}^2,&\max\{|T_{\alpha_k}|, |T_{\beta_k}|\} < \lambda_n,
\end{array}\right.
\end{eqnarray}
where  $P^{(k)}_{JS}$
is given in (\ref{Eq-4-5}), the value of $\lambda_n$ is equivalent to that of $P_{AJS}$ in (\ref{Eq3-2}).
For controlling the FWER,
we can reject $H^{(k)}_0$ if the AJS test statistic $P^{(k)}_{AJS}$ is much smaller than $\delta/d$. In other words, an estimated index set of significant mediators with AJS test is
$\hat{\Omega}_{AJS} = \{k: P^{(k)}_{AJS} < \delta/d, k=1,\cdots,d\}$, where $P^{(k)}_{AJS}$ is defined in (\ref{Eq-4-6}).


\begin{theorem}\label{Th2}
Under the significance level $\delta$,  {with probability approaching one}, the FWER of the AJS test for (\ref{Eq-4-1})  asymptotically satisfies
\begin{eqnarray}\label{MU-JS48}
\lim\limits_{n\rightarrow \infty }FWER(AJS) \leq \delta.
\end{eqnarray}
\end{theorem}
\begin{remark}
The $FWER(AJS)$ is asymptotically controlled below the significance level $\delta$ for small-scale multiple testing. We emphasize that the AJS cannot be directly applied to high-dimensional mediators.   In the context of large-scale mediators, advanced high-dimensional statistical techniques, such as employing variable screening methods for mediator dimension reduction, are imperative, which falls beyond the scope of this paper. 
\end{remark}
 
\begin{remark}
The ASobel method can be extended in a similar manner to address the multiple testing problem presented in (\ref{Eq-4-1}). As discussed in section \ref{sec-2}, AJS outperforms ASobel numerically, hence this topic is not explored further in this paper.
\end{remark}

\section{Adjusted Sobel-Type Confidence Intervals}\label{section-4-ASobel}
\setcounter{equation}{0}
In the current section, we focus on the estimation method for confidence intervals of mediation effects $\alpha_k\beta_k$'s, which plays a crucial role in comprehending the mediation mechanism with desirable levels of confidence. The index $k$ in $\alpha_k\beta_k$ is omitted for the sake of convenience, while maintaining the same level of generality.  The AJS method in Section \ref{sec-2} is proposed for conducting hypothesis testing, but it cannot be utilized for constructing confidence intervals of mediation effects. The Sobel-type (or normality-based) method is consistently employed in the literature to examine confidence intervals of mediation effects. To be specific, as $n\rightarrow \infty$, Sobel's method assumes that
\begin{eqnarray}\label{Sobel-CI-18}
\hat{\sigma}_{\alpha\beta}^{-1} ({\hat{\alpha}\hat{\beta} - \alpha\beta}) \stackrel{\mathcal{D}}{\longrightarrow} N(0,1),
\end{eqnarray}
where $\stackrel{\mathcal{D}}{\longrightarrow}$ denotes convergence in distribution, $\hat{\alpha}$ and $\hat{\beta}$ are the estimates for $\alpha$ and $\beta$, respectively; $\hat{\sigma}_{\alpha}$ and $\hat{\sigma}_{\beta}$ are the
estimated standard errors of $\hat{\alpha}$ and $\hat{\beta}$, respectively; $\hat{\sigma}_{\alpha\beta} = \{\hat{\alpha}^2\hat{\sigma}^2_{\beta} + \hat{\beta}^2\hat{\sigma}^2_{\alpha}\}^{1/2}$. The focus of our study lies in constructing $100(1-\delta)\%$ confidence intervals for mediation effects, where $\delta$ is commonly selected as $0.05$. Based on 
(\ref{Sobel-CI-18}), the Sobel-type $100(1-\delta)\%$  confidence interval for $\alpha\beta$ is given by
\begin{eqnarray}\label{CI-sobel-19}
{\rm CI_{Sobel}}=[\hat{\alpha}\hat{\beta} - N_{1-\delta/2}(0,1)\hat{\sigma}_{\alpha\beta},~\hat{\alpha}\hat{\beta} + N_{1-\delta/2}(0,1)\hat{\sigma}_{\alpha\beta}],
\end{eqnarray}
where $N_{1-\delta/2}(0,1)$ is the $(1-\delta/2)$-quantile of $N(0,1)$. However, the confidence interval ${\rm CI_{Sobel}}$ provided in (\ref{CI-sobel-19}) is excessively wide when both $\alpha$ and $\beta$ are equal to zero. e.g., the coverage probability of the $95\%$ confidence interval provided in (\ref{CI-sobel-19}) approaches unity.

The asymptotic distribution of $\hat{\sigma}_{\alpha\beta}^{-1} ({\hat{\alpha}\hat{\beta} - \alpha\beta})$ is $N(0,1/4)$ instead of $N(0,1)$ when $\alpha=\beta=0$, as stated in \cite{Liu-JASA-DACT-2022}. To amend the issue of ${\rm CI_{Sobel}}$ in the case of $\alpha=\beta=0$, we propose a novel adjusted Sobel-type confidence interval for $\alpha\beta$ as follows,
\begin{eqnarray}\label{CI-Asobel-20}
{\rm CI_{ASobel}}
&=&\left\{\begin{array}{ll} [\hat{\alpha}\hat{\beta} - N_{1-\delta/2}(0,1)\hat{\sigma}_{\alpha\beta},~\hat{\alpha}\hat{\beta} + N_{1-\delta/2}(0,1)\hat{\sigma}_{\alpha\beta}],& \max\{|T_\alpha|, |T_\beta|\} \geq \lambda_n,\nonumber\\
{[\hat{\alpha}\hat{\beta} - N_{1-\delta/2}(0,1/4)\hat{\sigma}_{\alpha\beta},~\hat{\alpha}\hat{\beta} + N_{1-\delta/2}(0,1/4)\hat{\sigma}_{\alpha\beta}]},&\max\{|T_\alpha|, |T_\beta|\} < \lambda_n,
\end{array}\right.\\
\end{eqnarray}
where $N_{1-\delta/2}(0,1/4)$ is the $(1-\delta/2)$-quantile of $N(0,1/4)$, $T_\alpha$ and $T_\beta$ are defined in (\ref{Eq2-3}), and  $\lambda_n =  \sqrt{n}/\log(n)$. The thresholding framework in (\ref{CI-Asobel-20}) shares a similar concept with the AJS and ASobel introduced in Section \ref{sec-2}.

\begin{theorem}\label{Th3}
As the sample size $n$ approaches infinity, {with probability approaching one}, the coverage probability of the $100(1-\delta)\%$ confidence interval given in (\ref{CI-Asobel-20}) asymptotically satisfies
\begin{eqnarray*}
\lim_{n\rightarrow\infty}\mathbb{P}(\alpha\beta \in {\rm CI_{ASobel}})
= 1-\delta.
\end{eqnarray*} 
\end{theorem}

The subsequent discussion presents a series of comparisons between ${\rm CI_{Sobel}}$ and  ${\rm CI_{ASobel}}$. The Sobel-type confidence interval can be derived directly to fulfill the following expression:
\begin{eqnarray*}
\lim_{n\rightarrow\infty}\mathbb{P}(\alpha\beta \in {\rm CI_{Sobel}})
&=&\left\{\begin{array}{ll} 2\Phi_{N(0,1/4)}(N_{1-\delta/2}(0,1))-1,& \alpha=\beta=0,\nonumber\\
1-\delta,&others,
\end{array}\right.\\
\end{eqnarray*}
where ${\rm CI_{Sobel}}$ is given in (\ref{CI-sobel-19}). By deducing the difference between $\lim_{n\rightarrow\infty}\mathbb{P}(\alpha\beta \in {\rm CI_{Sobel}})$ and $1-\delta$ under the case of $\alpha=\beta=0$, we have 
\begin{eqnarray*}
\lim_{n\rightarrow\infty}\mathbb{P}(\alpha\beta \in {\rm CI_{Sobel}}) - (1-\delta) &=& 2\Phi_{N(0,1/4)}(N_{1-\delta/2}(0,1))-2 + \delta\\
&>&0.
\end{eqnarray*}
Hence, the asymptotic coverage probability of ${\rm CI_{ASobel}}$ is much better than that of ${\rm CI_{Sobel}}$ under $\alpha=\beta=0$. Furthermore, the averaged length of ${\rm CI_{ASobel}}$ in the case of $\alpha=\beta=0$, $2N_{1-\delta/2}(0,1/4)\hat{\sigma}_{\alpha\beta}$ , is significantly shorter compared to that of  ${\rm CI_{Sobel}}$, which is $2N_{1-\delta/2}(0,1)\hat{\sigma}_{\alpha\beta}$. 
The performance of ${\rm CI_{Sobel}}$ and ${\rm CI_{ASobel}}$ will be compared through numerical simulations.


\section{Numerical Studies}\label{sec-5}
\setcounter{equation}{0}
\subsection{Size and Power of Single-Mediator Testing}\label{sec-5-1}
In this section, we conduct some simulations to evaluate the performance of the ASobel and AJS tests for  $H_0: \alpha\beta =0$ in the context of one mediator. Under the framework of mediation analysis, we consider three kinds of outcomes with one continuous mediator: {\it linear mediation model} (continuous outcome), {\it logistic mediation model} (binary outcome) and {\it Cox mediation model} (time-to-event outcome).  Specifically, {\red the mediator $M$ is generated from the linear model $M=\alpha X + \bfeta^\prime \mathbf{Z} + e$, where $X$ and $e$ follow from $N(0,1)$, $\mathbf{Z}=(Z_1,Z_2)^\prime$ with $Z_1$ and $Z_2$ being independent random variables following $N(0,1)$, $\bfeta = (0.5, 0.5)^\prime$.} The random outcomes are generated from the following three models:

$\bullet$ {\it Linear mediation model}:~ {\red $Y= \gamma X + \beta M + \btheta^\prime \mathbf{Z} + \epsilon$, where $\gamma=0.5$ and $\btheta = (0.5, 0.5)^\prime$.}

$\bullet$ {\it Logistic mediation model}: {\red Let $Y\in \{0,1\}$ be the binary outcome, and 
$${P}(Y=1|X, M, \mathbf{Z})= \frac{\exp(\gamma X + \beta M + \btheta^\prime \mathbf{Z})}{1+\exp(\gamma X + \beta M + \btheta^\prime \mathbf{Z})},$$ where $\gamma=0.5$ and $\btheta = (0.5, 0.5)^\prime$. }

$\bullet$ {\it Cox mediation model}: Let $T$ be the failure time, and $C$ be the censoring time. The observed survival time is $Y= \min(T,C)$. Following \cite{Cox1972}, the conditional hazard function of $T$ is {\red $\lambda(t|X, M, \mathbf{Z}) = \lambda_0(t)\exp(\gamma X + \beta M + \btheta^\prime \mathbf{Z})$}, where $\lambda_0(t) =1$ is the baseline hazard function, $\gamma=0.5$, $\btheta = (0.5, 0.5)^\prime$; $C$ is generated from $U(0,c_0)$ with $c_0$ being chosen as a number such that the censoring rate is about $30\%$.

\begin{table}[htp] 
  \begin{center}
    \caption{The size and power of hypothesis testing with linear mediation model$^\ddag$.}
    \label{tab:1}
    \vspace{0.1in} \small
    \begin{tabular}{llcccccccccc}
      \hline
      & $(\alpha, \beta)$ & &Sobel &JS &Bootstrap&ASobel &AJS \\
      \hline
    $n=200$
   & (0, 0)    & &0          & 0.0016 & 0.0010 & 0.0432 & 0.0460\\
   & (0, 0.5)   & &0.0406    &  0.0498 &0.0556 &0.0406 & 0.0498\\
  & (0.5, 0)   & &0.0420    & 0.0522 & 0.0568 & 0.0420 & 0.0522\\
  &(0.15, 0.15)& & 0.1190    &0.3068 & 0.2774 & 0.4184 &  0.5124\\
  &(0.25, 0.25)& &0.7470    & 0.8814 & 0.7584 &  0.7834 & 0.8978\\
      \hline
    $n=500$
   & (0, 0)    & &0         & 0.0034 & 0.0008 & 0.0482 & 0.0446\\
  & (0, 0.5)   & &0.0540    & 0.0568 & 0.0542 & 0.0540 &  0.0568\\
  & (0.5, 0)   & &0.0460    & 0.0496 & 0.0546 & 0.0460 & 0.0496\\
  &(0.15, 0.15)& &0.6820    & 0.8334 & 0.8218 & 0.8740 & 0.9090\\
  &(0.25, 0.25)& &0.9990    & 0.9998 & 0.9946 &  0.9990 &0.9998\\
 \hline
    $n=1000$
   & (0, 0)    & &0.0002    & 0.0026 & 0.0012 & 0.0498 & 0.0496\\
  & (0, 0.5)   & &0.0456    & 0.0482 & 0.0520 & 0.0456 & 0.0482\\
  & (0.5, 0)   & &0.0466    & 0.0476 & 0.0486 & 0.0466 & 0.0476\\
  &(0.15, 0.15)& &0.9886    & 0.9950 & 0.9952 & 0.9954 & 0.9974\\
  &(0.25, 0.25)& &1    & 1 & 1 & 1 & 1\\
       \hline
    \end{tabular}
  \end{center}
\end{table}

Under the regularity conditions in \cite{Vanderweele-SII-2009}, \cite{VanderWeele-AJE-2010} and \cite{VanderWeele2011-Epidemiology}, the product term $\alpha\beta$  can be interpreted as
the causal mediating effect of $M$ along the  pathway $X\rightarrow M \rightarrow Y$ for the  three mediation models (see Figure \ref{fig:1}). For comparison, we also use the traditional Sobel, JS and {\red Bootstrap}  methods for testing $H_0: \alpha\beta =0$, where the significance level is $\delta$ = 0.05. {\red The popular quantile Bootstrap test method is employed for assessing the mediation effect, and $H_0$ will be rejected if 0 does not fall within the interval $[Q_{\delta/2}(\{\hat{\alpha}^{(b)}\hat{\beta}^{(b)}\}_{b=1}^B), ~ Q_{1-\delta/2}(\{\hat{\alpha}^{(b)}\hat{\beta}^{(b)}\}_{b=1}^B)]$, where $Q_{\delta/2}(\cdot)$
is the $(\delta/2)$-empirical quantile function, $\hat{\alpha}^{(b)}$ and $\hat{\beta}^{(b)}$ are 
the corresponding parameter estimators with the $b$-th Bootstrap samples, $b=1,\cdots,B$. }

\begin{figure}[htp] 
  \centering
  \begin{subfigure}{0.4\textwidth}
    \includegraphics[width=\textwidth]{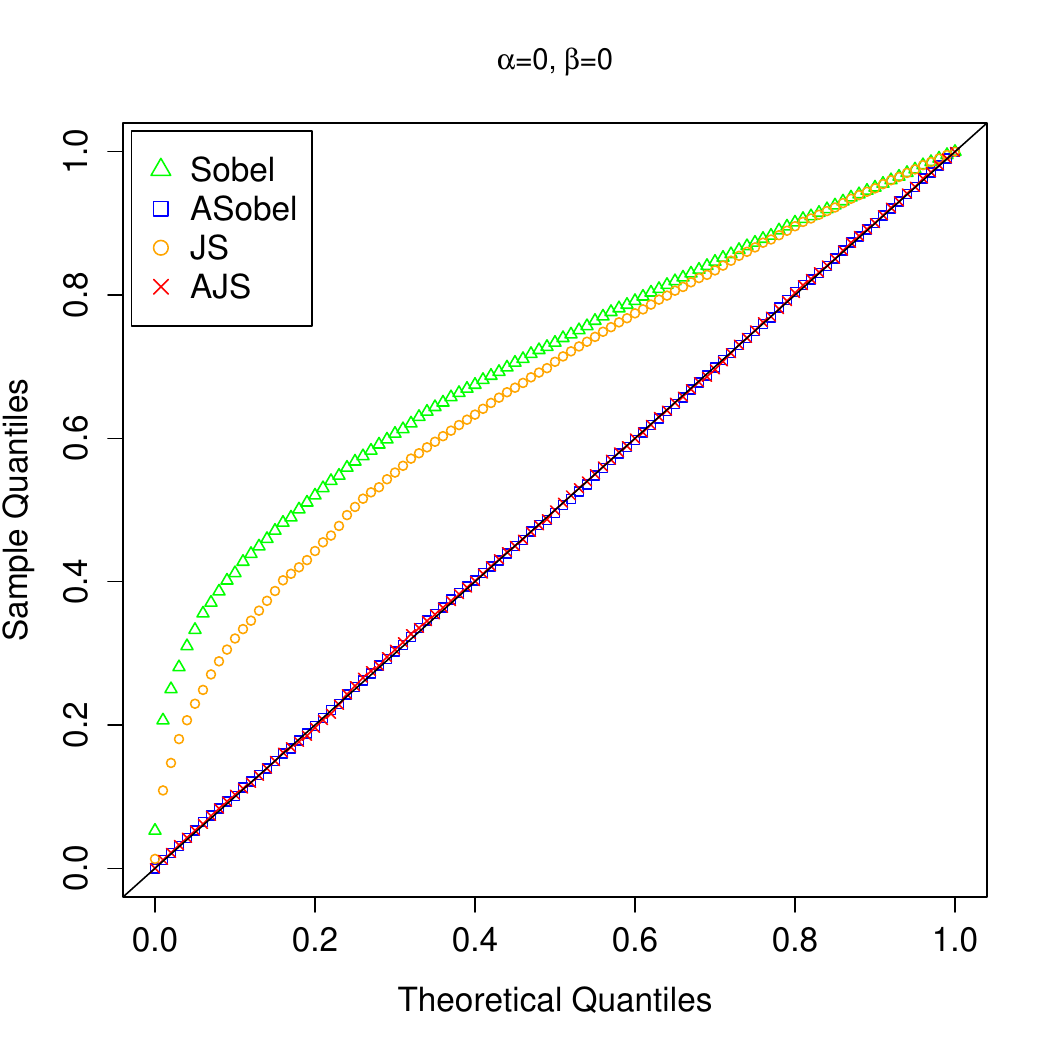}
    \caption{Q-Q plots of p-values under $H_{00}$.}
  \end{subfigure}
  \begin{subfigure}{0.4\textwidth}
    \includegraphics[width=\textwidth]{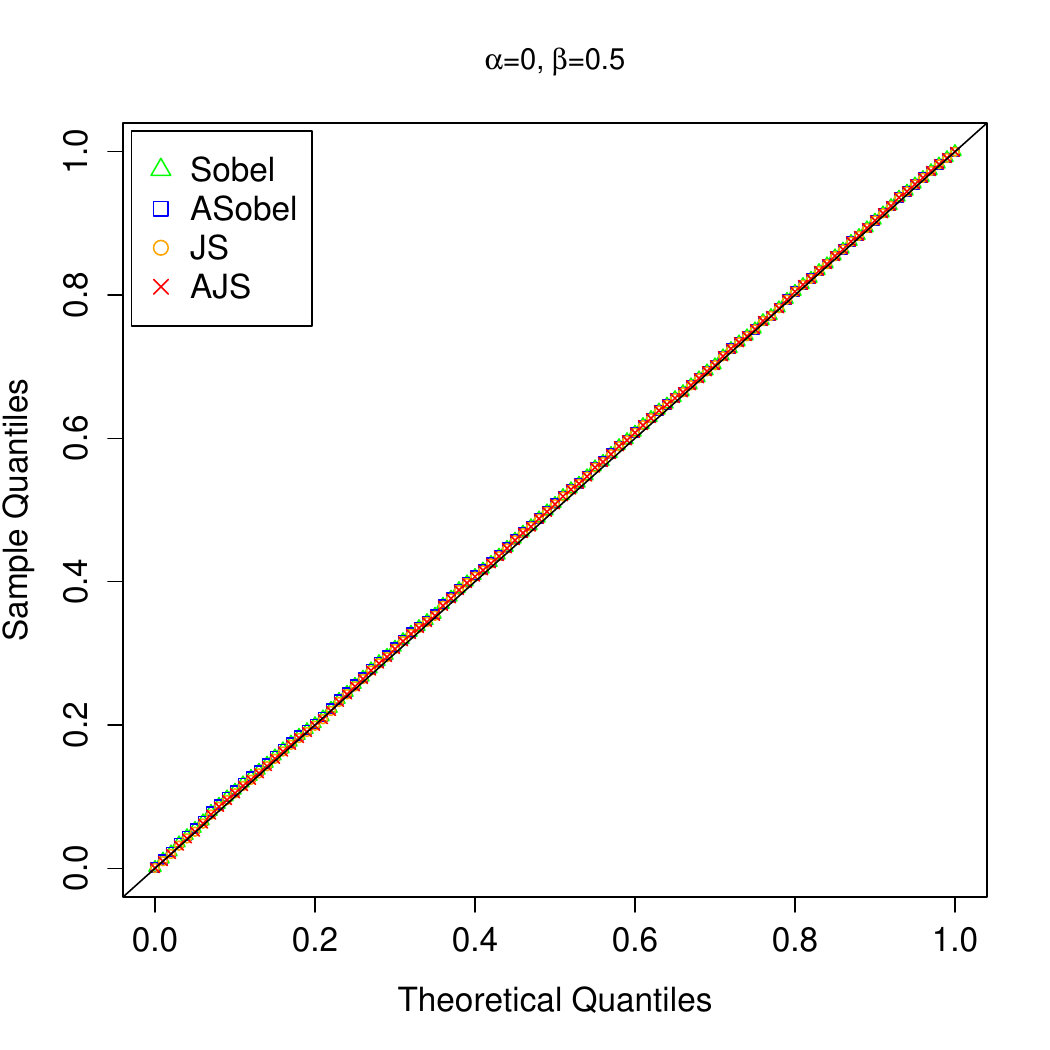}
    \caption{Q-Q plots of p-values under $H_{01}$.}
  \end{subfigure}
    \begin{subfigure}{0.4\textwidth}
    \includegraphics[width=\textwidth]{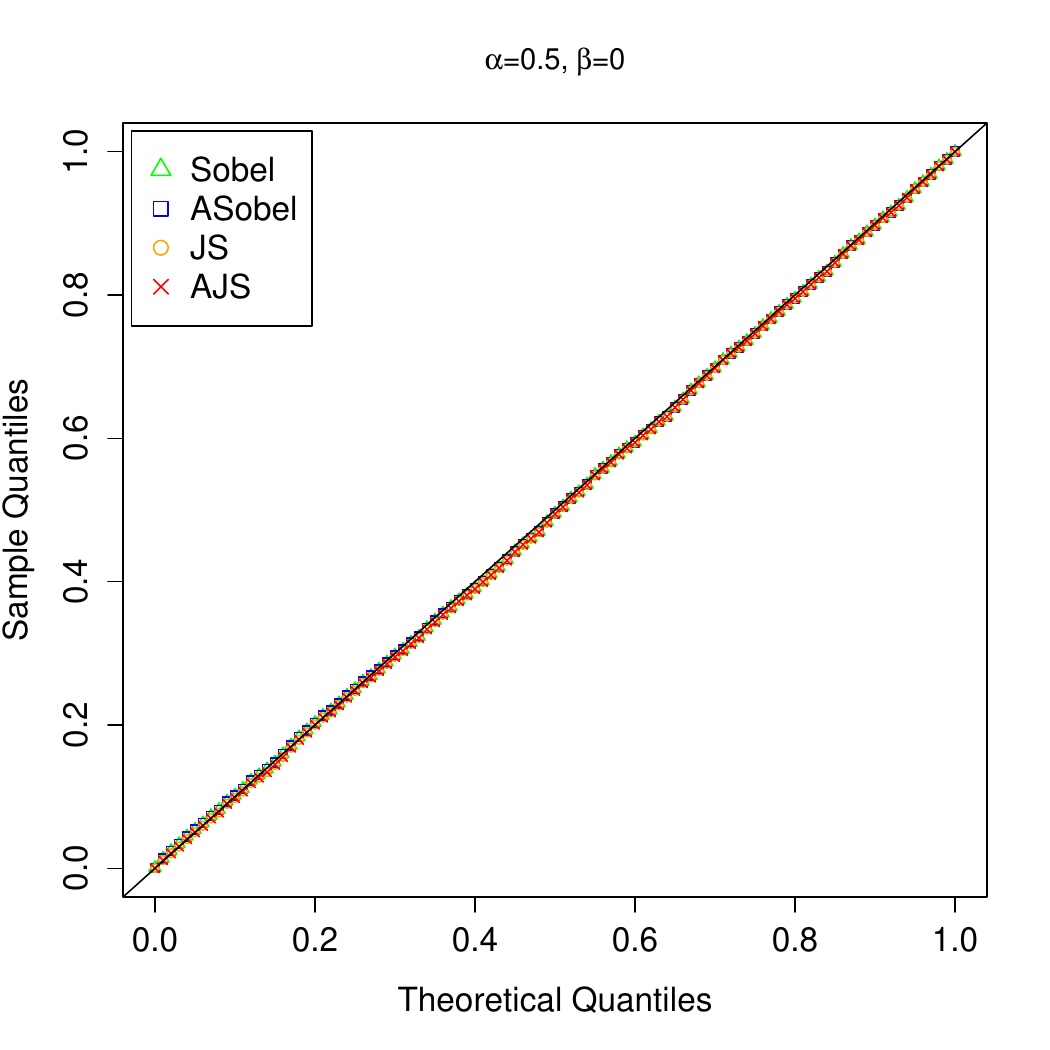}
    \caption{Q-Q plots of p-values under $H_{10}$.}
  \end{subfigure}
  \begin{subfigure}{0.4\textwidth}
    \includegraphics[width=\textwidth]{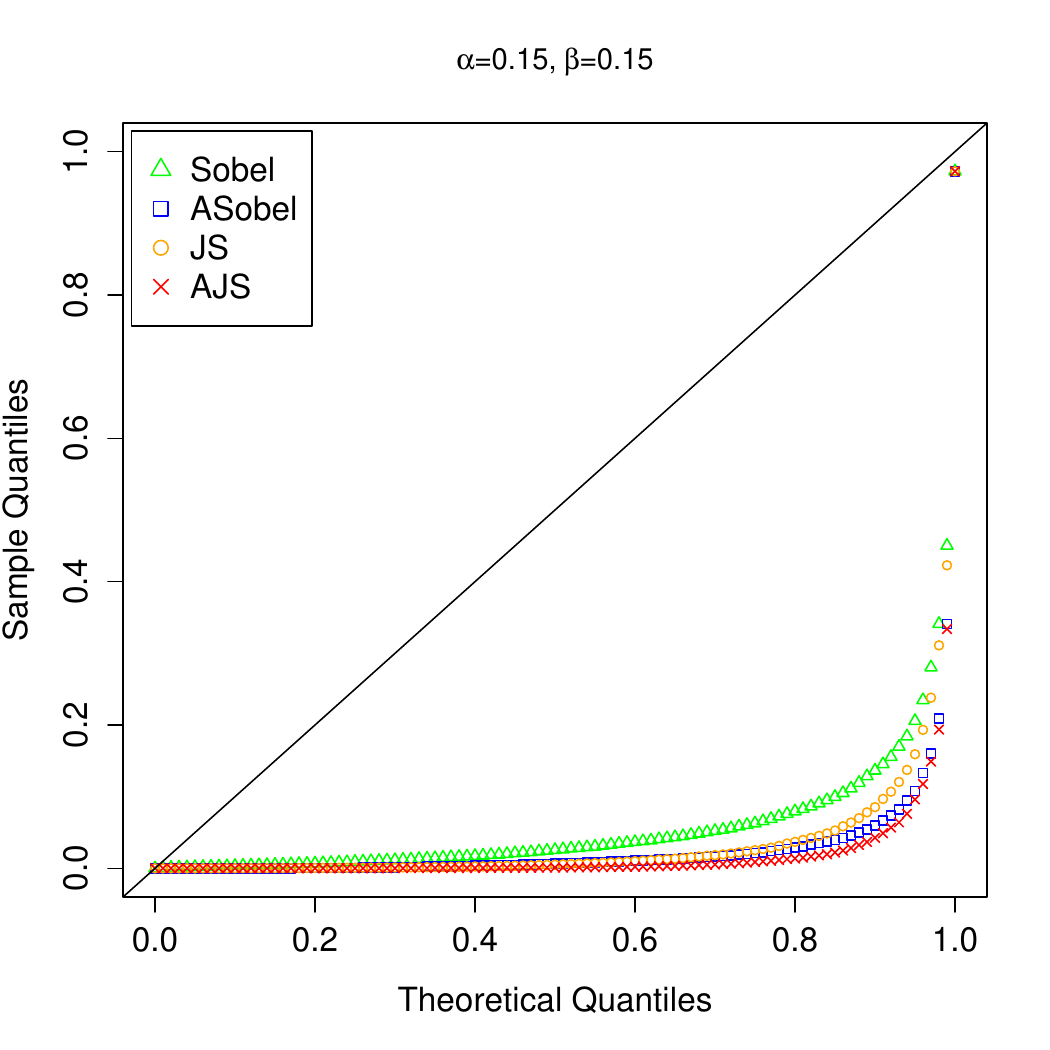}
    \caption{Q-Q plots of p-values under $H_{A}$.}
  \end{subfigure}
 \vspace{-0.1cm}
\begin{center}
  \caption{Q-Q plots of p-values under linear mediation model with $n = 500$.}
  \label{fig:6QQ}
\end{center}
\end{figure}

{\red All the simulation results are based on $5000$ repetitions, where $B=1000$ and the sample size is chosen as $n = 200$, $500$ and 1000, respectively.}
In Tables \ref{tab:1}, S.1 and S.2, we report the sizes and powers of the Sobel, JS, Bootstrap, ASobel and AJS methods when performing mediation tests with three kinds of outcomes. Under $\alpha=\beta=0$, the size of our proposed  AJS and ASobel are much better than those of JS, Sobel and Bootstrap methods, respectively. For small values of $\alpha$ and $\beta$, the power of  AJS is larger than that of ASobel and the conventional Sobel, JS and Bootstrap methods. These numerical findings are in line with the theoretical results of Theorem \ref{Th1}. In Figures \ref{fig:6QQ}, S.2 and S.3, we present the  Q-Q plots of p-values under linear, logistic and Cox  mediation models with $n = 500$. The Bootstrap method does not depend on the use of p-values when conducting hypothesis tests for $H_0$. The Q-Q plots  demonstrate that the Sobel, JS, ASobel, and AJS methods accurately approximate the distribution of their respective test statistics under either $H_{10}$ or $H_{01}$. The Sobel and JS tests, however, exhibit a conservative behavior, whereas the proposed ASobel and AJS tests still accurately approximate the distribution of their corresponding test statistics. The quantiles of p-values under $H_A$ increase in the following order: AJS, ASobel, JS, Sobel. The aforementioned finding is consistent with the power performance  presented in Table \ref{tab:1}.

\subsection{FWER and Power of Multiple-Mediators Testing}\label{sec5-2}

In this section,  we investigate the  performance of the AJS method when performing small-scale multiple testing for mediation effects via numerical simulations. The dimension of mediators is chosen as $d$ = 10, 15 and 20, respectively. Similar to Section \ref{sec-5-1}, we consider three kinds of outcomes in the context of multiple mediators.\\
$\bullet$ {\it Linear mediation model}:~ $Y= \gamma X + \bbeta^\prime \mathbf{M} +\btheta^\prime \mathbf{Z} + \epsilon$,
where $X$ and $\epsilon$ follow from $N(0,1)$, {\red $\mathbf{Z}=(Z_1,Z_2)^\prime$ with $Z_1$ and $Z_2$ being independent random variables following $N(0,1)$,}  $\gamma = 0.5$, $\btheta = (0.5, 0.5)^\prime$; $\mathbf{M}= (M_1,\cdots,M_d)^\prime$ is generated from a series of linear models $M_k = \alpha_k X + \bfeta_k^\prime \mathbf{Z} +e_k$. Here $\bfeta_k = (0.5, 0.5)^\prime$, and
$\mathbf{e} = (e_1,\cdots,e_d)^\prime$ is a multivariate normal vector with mean zero and
covariance matrix $\Sigma = (0.25^{|i-j|})_{i,j}$. The  parameter's settings are 
\begin{eqnarray*}
\balpha &=& (0.15, 0.05, 0.15, 0.15, 0.05, 0.5, 0.5, 0,\cdots, 0)^\prime,\\
\bbeta &=& (0.15, 0.05, 0.15,  0.05, 0.1, 0, 0, 0.5, 0.5, 0, \cdots, 0)^\prime.
\end{eqnarray*}
$\bullet$ {\it Logistic mediation model}:~ Let $Y\in \{0, 1\}$ be the binary outcome, and
$${P}(Y=1|X, \mathbf{M}, \mathbf{Z})= \frac{\exp(\gamma X + \bbeta^\prime \mathbf{M} + \btheta^\prime \mathbf{Z})}{1+\exp(\gamma X + \bbeta^\prime \mathbf{M} + \btheta^\prime \mathbf{Z})},$$
 where the variables $X$, $\mathbf{M}$, $\mathbf{Z}$ remain consistent with those in those of the linear mediation model, except for the settings for model parameters,
\begin{eqnarray*}
\balpha &=& (0.15, 0.2, 0.25, 0.5, 0.25, 0.65,  0, \cdots, 0)^\prime,\\
\bbeta &=& (0.25, 0.3, 0.35, 0.65, 0.55, 0, 0.55,  0,\cdots, 0)^\prime.
\end{eqnarray*}
$\bullet$ {\it Cox mediation model}: Given the covariate $X$, the mediator vector $\mathbf{M}$ and {\red the covariates vector $\mathbf{Z}$},  the Cox's conditional hazard function of failure time $T$ is $$\lambda(t|X, \mathbf{M}, \mathbf{Z}) = \lambda_0(t)\exp(\gamma X + \bbeta^\prime \mathbf{M} + \btheta^\prime \mathbf{Z}),$$
 where $\lambda_0(t) =1$, $\gamma = 0.5$, $X$, $\mathbf{M}$ and $\mathbf{Z}$ are generated in the same way as the  linear mediation model. We generate the censoring time $C$ from $U(0, c_0)$, where $c_0$  is being chosen  such that the censoring rate is about $30\%$. The settings of the model's parameters are
\begin{eqnarray*}
\balpha &=& (0.2, 0.35, 0.25, 0.15, 0.15, 0.5, 0, \cdots, 0)^\prime,\\
 \bbeta &=& (0.15, 0.25, 0.3, 0.15, 0.1, 0, 0.5,  0, \cdots, 0)^\prime.
\end{eqnarray*}

The index set of significant mediators is $\Omega = \{1,2,3,4,5\}$.
Based on \citeauthor{Vanderweele-SII-2009} (\citeyear{Vanderweele-SII-2009}; \citeyear{VanderWeele-AJE-2010}) and \cite{Huang-Yang2017}, the product term $\alpha_k\beta_k$ describes the causal mediation effect along the $k$th pathway $X \rightarrow M_k \rightarrow Y$ (see Figure \ref{fig:5}), where $k=1,\cdots,d$.  Under the significance level $\delta$, we consider the multiple testing problem $H_{0k}:~ \alpha_k \beta_k =0$, $k=1,\cdots,d$. The proposed AJS is compared with Sobel, JS,  JT\_Comp \cite[]{Comp-Huang}, DACT \cite[]{Liu-JASA-DACT-2022} and Bootstrap in terms of empirical FWER and Power. {\red The Bootstrap test method will reject $H_{0k}$ if 0 does not fall within the Bonferroni corrected interval $[Q_{\delta/(2d)}(\{\hat{\alpha}_k^{(b)}\hat{\beta}_k^{(b)}\}_{b=1}^B), ~ Q_{1-\delta/(2d)}(\{\hat{\alpha}_k^{(b)}\hat{\beta}_k^{(b)}\}_{b=1}^B)]$, where $\hat{\alpha}_k^{(b)}$ and $\hat{\beta}_k^{(b)}$ are 
the corresponding parameter estimators with the $b$-th Bootstrap samples, $b=1,\cdots,B$.}
The HDMT \cite[]{Dai2022-HDMT-JASA} is not applicable in our simulation scenarios, as it specifically targets high-dimensional mediation hypotheses with $d\rightarrow \infty$. 
All the results are based on 5000 repetitions, {\red where $B=1000$, the sample size is $n=200$, 500 and 1000, respectively.}

In Tables \ref{tab:2}, S.3 and S.4, we report the FWERs and Powers of {\red six test methods} in the context of linear, logistic and Cox mediation models, where the significance level is $\delta=$  0.05. The results demonstrate that the FWER and Power of AJS outperform the other five methods significantly. The powers of all methods are observed to increase with the sample size, while they decline with the increase in dimension $d$.  One explanation for this phenomenon is due to the fact that the estimated variances of model parameters are becoming larger as the increase of mediator's dimension under fixed sample size. 

\begin{table}[htp] 
  \begin{center}
    \caption{The FWER and power of multiple testing with linear mediation model.}
    \label{tab:2}
    \vspace{0.1in} \small
    \begin{tabular}{llcccccccccc}
      \hline
  & Dimension & &Sobel &JS & JT\_Comp & DACT &Bootstrap & AJS\\
      \hline
     $n=200$
   & $d=10$ &FWER &0.0062 &0.0192  & 0.1706 & 0.0116  & 0.0294 & 0.0232\\
      &     &Power&0.0011 &0.0195  & 0.0032 & 0.0146  & 0.0127 & 0.0764\\
   & $d=15$ &FWER &0.0052 &0.0126  & 0.1982 & 0.0252  & 0.0176 & 0.0282\\
      &     &Power&0.0004 &0.0147  & 0.0034 & 0.0236  & 0.0092 & 0.0568\\
   & $d=20$ &FWER &0.0038 &0.0112  & 0.2176 & 0.0486  & 0.0186 & 0.0338\\
      &     &Power&0.0002 &0.0103  & 0.0042 & 0.0314  & 0.0076 & 0.0458\\
      \hline
     $n=500$
   & $d=10$ &FWER &0.0166 &0.0222  & 0.2074 & 0       & 0.0296 & 0.0262\\\
      &     &Power&0.0452 &0.1836  & 0.0099 & 0.0252  & 0.1265 & 0.3286\\
   & $d=15$ &FWER &0.0080 &0.0134  & 0.2282 & 0.0006  & 0.0230 & 0.0306\\
      &     &Power&0.0299 &0.1560  & 0.0135 & 0.0646  & 0.1094 & 0.2954\\
   & $d=20$ &FWER &0.0084 &0.0108  & 0.2748 & 0.0086  & 0.0190 & 0.0388\\
      &     &Power&0.0216 &0.1387  & 0.0199 & 0.1160  & 0.0965 & 0.2762\\
      \hline
     $n=1000$
   & $d=10$ &FWER &0.0172 &0.0198  & 0.2102 & 0       & 0.0268 & 0.0246\\
      &     &Power&0.3055 &0.4042  & 0.0505 & 0.0279  & 0.3547 & 0.5211\\
   & $d=15$ &FWER &0.0100 &0.0128  & 0.2382 & 0.0002  & 0.0164 & 0.0342\\
      &     &Power&0.2669 &0.3864  & 0.0605 & 0.1408  & 0.3390 & 0.5004\\
   & $d=20$ &FWER &0.0086 &0.0100  & 0.2748 & 0.0020  & 0.0178 & 0.0352\\
      &     &Power&0.2361 &0.3734  & 0.0771 & 0.2534  & 0.3219 & 0.4857\\
      \hline
    \end{tabular}
  \end{center}
\end{table}

%

\begin{table}[htp] 
  \begin{center}
    \caption{The coverage probability and length of 95\% confidence interval with linear mediation model$^\ddag$.}
    \label{tab:4-CI}
    \vspace{0.1in} \small
    \begin{tabular}{llcccccccccccc}
      \hline
   &&&\multicolumn{3}{c}{CP} &&   \multicolumn{3}{c}{LCI} \\
      \cline{4-6}\cline{8-10}
      &$(\alpha_k, \beta_k)$ & &Sobel & Bootstrap & ASobel & & Sobel & Bootstrap&ASobel \\
      \hline
   $n=200$&
   (0, 0)      & &1      & 0.9986 & 0.9482 &&0.02665 & 0.04009 & 0.01388   \\
& (0.35, 0)    & &0.9676 & 0.9454 & 0.9632 &&0.10811 & 0.11234 & 0.10764   \\
& (0.5, 0)     & &0.9522 & 0.9460 & 0.9522 &&0.15321 & 0.15638 & 0.15321   \\
&(0, 0.35)     & &0.9706 & 0.9394 & 0.9614 &&0.10098 & 0.11246 & 0.10009   \\
&(0, 0.5)      & &0.9614 & 0.9414 & 0.9614 &&0.14158 & 0.15478 & 0.14158   \\
&(0.25, 0.35)  & &0.9376 & 0.9466 & 0.9376 &&0.12626 & 0.13371 & 0.12606   \\
&(0.45, 0.5)   & &0.9424 & 0.9450 & 0.9424 &&0.19413 & 0.20249 & 0.19413   \\
      \hline
   $n=500$&
   (0, 0)      & &0.9996 & 0.9972 & 0.9548 &&0.01024 & 0.01558 & 0.00513   \\
& (0.35, 0)    & &0.9572 & 0.9474 & 0.9572 &&0.06678 & 0.06769 & 0.06678   \\
& (0.5, 0)     & &0.9520 & 0.9472 & 0.9520 &&0.09488 & 0.09527 & 0.09488   \\
&(0, 0.35)     & &0.9604 & 0.9498 & 0.9604 &&0.06238 & 0.06834 & 0.06237   \\
&(0, 0.5)      & &0.9498 & 0.9440 & 0.9498 &&0.08817 & 0.09578 & 0.08817   \\
&(0.25, 0.35)  & &0.9474 & 0.9484 & 0.9474 &&0.07813 & 0.08223 & 0.07813   \\
&(0.45, 0.5)   & &0.9424 & 0.9410 & 0.9424 &&0.12088 & 0.12599 & 0.12088   \\
      \hline
   $n=1000$&
   (0, 0)      & &0.9996 & 0.9986 & 0.9472 &&0.00510 & 0.00776 & 0.00255   \\
& (0.35, 0)    & &0.9534 & 0.9454 & 0.9534 &&0.04668 & 0.04684 & 0.04668   \\
& (0.5, 0)     & &0.9488 & 0.9468 & 0.9488 &&0.06658 & 0.06658 & 0.06658   \\
&(0, 0.35)     & &0.9530 & 0.9484 & 0.9530 &&0.04376 & 0.04763 & 0.04376   \\
&(0, 0.5)      & &0.9480 & 0.9484 & 0.9480 &&0.06228 & 0.06755 & 0.06228   \\
&(0.25, 0.35)  & &0.9462 & 0.9476 & 0.9462 &&0.05483 & 0.05762 & 0.05483   \\
&(0.45, 0.5)   & &0.9544 & 0.9534 & 0.9544 &&0.08502 & 0.08845 & 0.08502   \\
\hline
    \end{tabular}
  \end{center}
  {\vspace{0.0cm} \hspace{0.0cm}\footnotesize $\ddag$ ``CP" denotes the empirical coverage probability; ``LCI" denotes the length of 95\% confidence interval; ``Sobel" denotes the ${\rm CI_{Sobel}}$ in (\ref{CI-sobel-19});  ``ASobel" denotes the proposed ${\rm CI_{ASobel}}$ in (\ref{CI-Asobel-20}); ``Bootstrap" denotes the ${\rm CI_{Bootstrap}}$ in (\ref{CI-boot})}.
\end{table}

\subsection{Coverage Probability of Confidence Interval}

The performance of the ASobel-type confidence interval presented in Section \ref{section-4-ASobel} is evaluated through simulations conducted in this subsection. The {\red Bootstrap confidence interval}  and traditional Sobel-type confidence interval, ${\rm CI_{Sobel}}$, given in (\ref{CI-sobel-19}), are also considered for comparison. {\red The $100(1-\delta)\%$ Bootstrap confidence interval of mediation effect is given as
\begin{eqnarray}\label{CI-boot}
{\rm CI_{Bootstrap}} = [Q_{\delta/2}(\{\hat{\alpha}^{(b)}\hat{\beta}^{(b)}\}_{b=1}^B), ~ Q_{1-\delta/2}(\{\hat{\alpha}^{(b)}\hat{\beta}^{(b)}\}_{b=1}^B)],
\end{eqnarray}
where $\hat{\alpha}^{(b)}$ and $\hat{\beta}^{(b)}$ are 
the corresponding parameter estimators with the $b$-th Bootstrap samples, $b=1,\cdots,B$.}

 The data are generated in a similar manner as those described in Section \ref{sec5-2}, with the parameters chosen as (i) Linear mediation model : $\balpha = ( 0, 0.35, 0.5, 0, 0, 0.25, 0.45)^\prime$ and $\bbeta = (0, 0, 0, 0.35, 0.5, 0.35, 0.5)^\prime$; (ii) Logistic mediation model: $\balpha = (0, 0.35, 0.5, 0, 0, 0.35, 0.45)^\prime$ and $\bbeta = (0, 0, 0, 0.75, 0.8, 0.75, 0.85)^\prime$; (iii) Cox mediation model: $\balpha = (0, 0.35, 0.5, 0, 0, 0.45, 0.35)^\prime$ and $\bbeta = (0, 0, 0, 0.5, 0.45, 0.55, 0.35)^\prime$. {\red All the results are based on 5000 repetitions, where $B=1000$, the sample sizes are chosen as $n=200$, 500 and 1000, respectively. }

The coverage probability (CP) and length of the 95\% confidence interval (LCI) with ${\rm CI_{Sobel}}$, ${\rm CI_{Bootstrap}}$ and ${\rm CI_{ASobel}}$  are reported in Tables \ref{tab:4-CI}, S.5 and S.6.  The ${\rm CI_{ASobel}}$ outperforms ${\rm CI_{Sobel}}$  in terms of CP under $\alpha=\beta=0$. Additionally, the LCI of  ${\rm CI_{ASobel}}$ is significantly shorter compared to that of ${\rm CI_{Sobel}}$ when $\alpha\beta=0$. These findings are consistent  with the result of Theorem \ref{Th3}. {\red Lastly, the ASobel confidence interval is much better than Bootstrap method in terms of both CP and LCI in the simulations.}


\subsection{A comparison between AJS and \cite{AB-JRSSB-2023}'s method}

In this section, we conduct a simulation to compare the AJS with the adaptive bootstrap for JS test (AB-JS) of \cite{AB-JRSSB-2023}. The numerical experiments conducted by \cite{AB-JRSSB-2023} did not account for binary and survival outcomes with continuous mediators. Therefore, our focus is solely on linear mediation models. For the purpose of ensuring a fair comparison, we have adopted the same model settings as those presented in section 4 of \cite{AB-JRSSB-2023}. To be specific, we generate random data with the following models:
\begin{eqnarray*}
Y&=& c+\gamma X + \beta{M}+ \theta_1{Z_1} + \theta_2{Z_2} + \epsilon,\nonumber\\
M&=& c_m+\alpha X + \eta_1 Z_1 + \eta_2 Z_2 + e,
\end{eqnarray*}
where the exposure variable $X$ is generated from a Bernoulli distribution with a success probability of 0.5; the covariate $Z_1$ is  simulated from  $N (0, 1)$; the covariate $Z_2$ is simulated from the Bernoulli distribution with the success probability 0.5; two error terms
$\epsilon$ and $e$ are simulated independently from $N(0, 0.25)$; We set the parameters $(c, \theta_1, \theta_2) = (1,1,1)$,
$(c_m, \eta_1, \eta_2) = (1,1,1)$, and $\gamma=1$. We use the AJS and AB-JS to test the hypothesis
$H_0: \alpha \beta =0$, where the significance level is $\delta$ = 0.05. The AB-JS is implemented using the codes provided in the R package  {\tt ABtest}, which is publicly available at \url{https://github.com/yinqiuhe/ABtest }. The bootstrap number of AB-JS is set to 500, which is also selected for the simulation in \cite{AB-JRSSB-2023}. Additionally, in order to ensure a fair comparison between AJS and AB-JS, the threshold parameter has been selected as $\lambda_n = \sqrt{n}/\log(n)$ for both methods.

In Table \ref{tab:55-AB-JS}, we report the size and power of AJS and AB-JS tests, where the results are based on 1000 repetitions. It can be seen from Table \ref{tab:55-AB-JS} that the power of AB-JS is larger than AJS under $n=500$ and 1000. However, the computation speed of our AJS method significantly outperforms AB-JS due to the absence of a resampling procedure, which is inherent in the bootstrap-based approach utilized by AB-JS. The AJS and AB-JS exhibit comparable performance in terms of size and power for $n=3000$. In other words, the AJS exhibits comparable statistical efficiency to that of AB-JS when dealing with larger datasets.

\begin{table}[htp] 
  \begin{center}
    \caption{The comparison between AJS and AB-JS in terms of size and power$^\ddag$.}
    \label{tab:55-AB-JS}
    \vspace{0.1in} \small
    \begin{tabular}{llccccccccccccc}
      \hline
       &  &\multicolumn{3}{c}{$n=500$} &  & \multicolumn{3}{c}{$n=1000$} &  & \multicolumn{3}{c}{$n=3000$}\\
      \cline{3-5}\cline{7-9}\cline{11-13}
      $(\alpha, \beta)$ & &AJS & &AB-JS & & AJS & &AB-JS & & AJS & &AB-JS\\
      \hline
    
   (0, 0)  & &0.031 &  & 0.056  &&0.055 &  & 0.056   &&0.051 &  & 0.062  \\
  (0.5, 0) & &0.048 &  & 0.048   &&0.047 &  & 0.046   &&0.055 &  &  0.051 \\
(0, 0.5) & &0.046 &  & 0.054   &&0.045 &  & 0.045  &&0.047 &  & 0.052  \\
(0.15, 0.15)& &0.375 &  & 0.891   &&0.663 &  &  0.998  &&0.983 &  & 1 \\
(0.25, 0.25) & &0.797 &  & 1   &&0.978 &  & 1   &&1 &  & 1 \\
      \hline
    \end{tabular}
  \end{center}
  {\vspace{0.0cm} \hspace{0.0cm}\footnotesize $\ddag$  ``AJS" denotes our adjusted JS method; ``AB-JS" refers to the adaptive bootstrap JS test \cite[]{AB-JRSSB-2023}.}
\end{table}

\section{Real Data Examples}\label{sec-6}
\setcounter{equation}{0}
In this section, we apply our proposed AJS method for testing the mediation effects towards three real-world datasets with continuous, binary and time-to-event outcomes, respectively. The details about the three datasets and mediation analysing procedures are presented as follows:

{\bf Dataset I:} ({\it  continuous outcomes}). The Louisiana State University Health Sciences Center has explored the relationship between children weight and  behavior  through a survey of children, teachers and parents in Grenada. The dataset is publicly available within the R package {\tt mma}. To perform mediation analysis as that of \cite{mma-2017}, we set gender as the exposure $X$ (Male =0; Female = 1),  and the outcome $Y$ is body mass index (BMI). We consider three mediators: $M_1$ (join in a sport team or not), $M_2$ (number of hours of exercises per week) and $M_3$ (number of hours of sweating activities per week). Furthermore, there are three covariates $Z_1$ (age), $Z_2$ (number of people in family) and $Z_3$ (the number of cars in family). After
removing those individuals with missing data, we totally have 646 samples  when conducting mediation analysis in the context of {\red linear mediation model.}  We consider the linear mediation model to fit this dataset:
\begin{eqnarray*}
Y&=& c+\gamma X + \bbeta^\prime \mathbf{M}+ \btheta^\prime \mathbf{Z} + \epsilon,\nonumber\\
M_k&=& c_k+\alpha_k X + \bfeta_k^\prime \mathbf{Z }+ e_k,~~ k=1,2,3,
\end{eqnarray*}
where $Y$ is the continuous outcome, $\mathbf{M} = (M_1, M_2, M_3)^\prime$ is the vector of mediators, $\mathbf{Z} = (Z_1, Z_2, Z_3)^\prime$ is the vector of covariates. By \cite{VanderWeele-AJE-2010}, the product term $\alpha_k\beta_k$ can be interpreted as the causal mediating effect of $M_k$ along the pathway $X \rightarrow M_k \rightarrow Y$. Here we consider the multiple testing problem $H_{0}^{(k)}:~\alpha_k\beta_k =0$, $k=1,2,3$. The details of $P_{Sobel}$, $P_{ASobel}$$P_{JS}$ and $P_{AJS}$ are presented in Table \ref{tab:5}. The estimators $\hat{\alpha}_k$'s and $\hat{\beta}_k$'s along with their standard errors are also given in  Table \ref{tab:5}.  Particularly,
$P_{AJS} \leq P_{JS}$ and $P_{ASobel} \leq P_{Sobel}$ demonstrate that the proposed
 AJS and ASobel are superior to traditional  JS and Sobel, respectively. {\red In Table \ref{tab:CI5},
 we give the 95\% confidence intervals for mediation effects $\alpha_k\beta_k$'s, where ${\rm CI_{Sobel}}$, ${\rm CI_{ASobel}}$ and  ${\rm CI_{Bootstrap}}$ are defined in (\ref{CI-sobel-19}), (\ref{CI-Asobel-20}) and (\ref{CI-boot}), respectively. The Bootstrap confidence  ${\rm CI_{Bootstrap}}$ is calculated with $B=1000$.  The results from Table \ref{tab:CI5} demonstrate that the proposed adjusted Sobel-type method yields a significantly shorter and more reliable confidence interval compared to the conventional Sobel and Bootstrap method.}\\

{\bf Dataset II:} ({\it binary outcomes}).  The Job Search Intervention Study (JOBS II) is a randomized field experiment that investigates the efficacy of a job
training intervention on unemployed workers.  The dataset is publicly available within the R package {\tt mediation}.  Our research aims to investigate whether the workshop enhances future employment prospects by increasing job-search self-efficacy levels. To be specific, we study the mediating role of job-search self-efficacy between job-skills workshop
 and employment status. For this aim, 
we set the exposure $X$ as an indicator variable for whether participant was randomly
selected for the JOBS II training program (1 = assignment to
participation); the mediator $M_1$ is a continuous scale measuring the level of job-search
self-efficacy; the outcome $Y$ is a binary measure of employment (1 = employed). Furthermore, there are 9 covariates: $Z_1$ (age), $Z_2$ (sex; 1 = female),  $Z_3$ (level of economic hardship pre-treatment), $Z_4$ (measure of depressive symptoms pre-treatment),
  $Z_5$ (factor with seven categories for various occupations), $Z_6$ (factor with five categories for marital status), $Z_7$ (indicator variable for race; 1 = nonwhite), 
$Z_8$ (factor with five categories for educational attainment), $Z_9$ (factor with five categories for level of income).
After excluding individuals with missing data, we have a total of 899 samples for conducting mediation analysis within the framework of logistic mediation models:
\begin{eqnarray*}
\mathbb{P}({Y}=1|X, {M}_1, \mathbf{Z})&=&  \frac{\exp(c+\gamma X + \beta_1 {M_1}+ \btheta^\prime \mathbf{Z})}{1 + \exp(c+\gamma X + \beta_1 {M_1} + \btheta^\prime \mathbf{Z})},\\
M_1&=& c_1 + \alpha_1 X + \bfeta^\prime \mathbf{Z }+ e,
\end{eqnarray*}
where $Y\in \{0, 1\}$ is the binary outcome, $M_1$ is the  mediator, $\mathbf{Z} = (Z_1, \cdots, Z_9)^\prime$ is the vector of covariates. By \cite{VanderWeele-AJE-2010}, the product term $\alpha_1\beta_1$ can be interpreted as the causal mediating effect of $M_1$ along the pathway $X \rightarrow M_1 \rightarrow Y$. Here we consider the mediation testing problem $H_{0}:~\alpha_1\beta_1 =0$. The details of $P_{Sobel}$, $P_{ASobel}$, $P_{JS}$ and $P_{AJS}$ are presented in Table \ref{tab:5}. The estimators $\hat{\alpha}_1$'s and $\hat{\beta}_1$'s along with their standard errors are also given in  Table \ref{tab:5}. It seems that the $M_1$ has a significant mediating role between exposure and outcome. Particularly,
$P_{AJS} \leq P_{JS}$ and $P_{ASobel} \leq P_{Sobel}$ demonstrate that the proposed
 method works well in practical application. The 95\% confidence interval of the mediation effect is presented in Table \ref{tab:CI5}, which yields a similar conclusion to that of the dataset I. \\

{\bf Dataset III:} ({\it time-to-event outcomes}). We apply our proposed method to a dataset from The Cancer Genome Atlas (TCGA) lung cancer cohort study, where the data are freely available at {https://xenabrowser.net/datapages/}. There are 593 patients with non-missing  clinical and epigenetic information. From \cite{zhang-surHIMA-2021}, we use seven DNA methylation markers as potential mediators: $M_1$ (cg02178957), $M_2$ (cg08108679), $M_3$ (cg21926276), $M_4$ (cg26387355), $M_5$ (cg24200525), $M_6$ (cg07690349) and $M_7$ (cg26478297). The exposure $X$ is defined as the number of packs smoked per years, and the survival time is the outcome variable. Two hundred forty three patients died during the follow-up, and the censoring  rate is  59\%. We are interested in testing the mediation effects of DNA methylation markers along the pathways from smoking
to survival of lung cancer patients. Four covariates are included: $Z_1$ (age at initial diagnosis), $Z_2$ (gender; male = 1, female=0), $Z_3$ (tumor stage; Stage I = 1, Stage II = 2, Stage III = 3, Stage IV = 4), and $Z_4$ (radiotherapy; yes = 1, no = 0). We use the following Cox mediation model to fit this dataset:
\begin{eqnarray*}
\lambda(t|X, \mathbf{M}, \mathbf{Z})&=& \lambda_0(t){\exp(\gamma X + \bbeta^\prime \mathbf{M}+ \btheta^\prime \mathbf{Z})}\\
M_k&=& c_k + \alpha_k X + \bfeta_k^\prime \mathbf{Z }+ e_k,~~ k=1,\cdots,7,
\end{eqnarray*}
where $\lambda_0(t)$ is the baseline hazard function, $\mathbf{M} = (M_1, \cdots, M_7)^\prime$ is the vector of mediators, $\mathbf{Z} = (Z_1, Z_2, Z_3, Z_4)^\prime$ is the vector of covariates, $e_k$'s are random errors. Based on \cite{Huang-Yang2017},
the term $\alpha_k\beta_k$ is the causal mediation effect of the $k$th mediator. We consider the multiple testing $H_{0}^{(k)}:~\alpha_k\beta_k=0$, $k=1, \cdots, 7$.  Table \ref{tab:5} presents the statistics $P_{Sobel}$,  $P_{JS}$ and $P_{AJS}$, along with parameter estimates and their standard errors. In view of the fact that  $P_{AJS} \leq P_{JS}$ and $P_{ASobel} \leq P_{Sobel}$, the proposed method is desirable when performing mediation analysis in practical applications.  The 95\% confidence intervals of the mediation effects are presented in Table \ref{tab:CI5}, which supports a similar conclusion as that derived from the dataset I.

\begin{table}[htp] 
  \begin{center}
    \caption{The p-values and parameter estimates in three real-world examples.}
    \label{tab:5}
    \vspace{0.1in} \small
    \begin{tabular}{cccccccccccc}
      \hline
    Datasets  & Mediators & &$P_{Sobel}$ &$P_{ASobel}$ &$P_{JS}$ &$P_{AJS}$ & $\hat{\alpha}_k$($\hat{\sigma}_{\alpha_k}$) &$\hat{\beta}_k$($\hat{\sigma}_{\beta_k}$)\\
      \hline
 I  & $M_1$ &&0.03333&0.00002 &0.00359&0.00001 &-0.1130 (0.0388) & -0.9822 (0.3150)\\
    & $M_2$ &&0.25023&0.02147 &0.11901&0.01416 &-0.1234 (0.0791) &0.2651 (0.1557)\\
    & $M_3$ &&0.26903&0.02706 &0.19525&0.03812 &0.1169 (0.0551) &0.2922 (0.2256)\\
 \hline
 II  & $M_1$ &&0.21183&0.01252 &0.11626&0.01352 &0.0774 (0.0493) &0.1356 (0.0659)\\
 \hline
  III& $M_1$ &&0.12043&0.00190 &0.02822&0.00080 &-0.0129 (0.0059) &1.2816 (0.5841)\\
     & $M_2$ &&0.02527&0.00001 &0.00542&0.00003 &-0.0092 (0.0024) &-2.8537 (1.0262)\\
     & $M_3$ &&0.10093&0.10093 &0.08030&0.08030 &-0.0094 (0.0054) &-3.4795 (0.7357)\\
     & $M_4$ &&0.10074&0.00103 &0.02692&0.00072 &-0.0125 (0.0051) &-1.4994 (0.6776)\\
     & $M_5$ &&0.15777&0.15777 &0.13013&0.13013 &-0.0033 (0.0022) &6.2711 (1.5944)\\
     & $M_6$ &&0.05167&0.00010&0.02247&0.00051 &-0.0162 (0.0071) &1.9535 (0.5246)\\
     & $M_7$ &&0.08990& 0.00069 &0.05725&0.00328 &-0.0256 (0.0068) &-0.8417 (0.4426)\\
    \hline
    \end{tabular}
  \end{center}
\end{table}

\begin{table}[htp] 
  \begin{center}
    \caption{The 95\% confidence intervals for mediation effects in three real-world examples.}
    \label{tab:CI5}
    \vspace{0.1in} \small
    \begin{tabular}{cccccccccccc}
      \hline
    Datasets  & Mediators & &${\rm CI_{Sobel}}$ &&${\rm CI_{Bootstrap}}$&&${\rm CI_{ASobel}}$&&$\hat{\alpha}_k\hat{\beta}_k$\\
      \hline
 I  & $M_1$ &&[0.0088, 0.2133]  &&[0.0234, 0.2251] && [0.0599, 0.1621]&& 0.1110\\
    & $M_2$ &&[-0.0885, 0.0231] &&[-0.1013, 0.0083]&& [-0.0606, -0.0048]&& -0.0327\\
    & $M_3$ &&[-0.0264,  0.0947] &&[-0.0189, 0.1022]&& [0.0039, 0.0644] &&0.0342\\
 \hline
 II  & $M_1$ &&[-0.0059, 0.0269] &&[-0.0022, 0.0335]&& [0.0023, 0.0187]&& 0.0105\\
 \hline
  III& $M_1$ &&[-0.0375, 0.0044] &&[-0.0459, 0.0003]&& [-0.0271, -0.0061]&& -0.0166\\
     & $M_2$ &&[0.0033, 0.0491]   &&[0.0015, 0.0533]&& [0.0147, 0.0377]& &0.0262\\
     & $M_3$ &&[-0.0064, 0.0715] &&[-0.0062, 0.0835]&& [-0.0064, 0.0715]&& 0.0326\\
     & $M_4$ &&[-0.0036, 0.0409] &&[-$10^{-5}$, 0.0471]&& [0.0075, 0.0298]&& 0.0187\\
     & $M_5$ &&[-0.0499, 0.0081] &&[-0.0574, 0.0086]&& [-0.0499, 0.0081] &&-0.0209\\
     & $M_6$ &&[-0.0635, 0.0002] &&[-0.0733, -0.0044]&& [-0.0476, -0.0157] &&-0.0316\\
     & $M_7$ &&[-0.0034, 0.0464] &&[$4\times 10^{-5}$, 0.0519]&& [0.0091, 0.0339]&& 0.0215\\
    \hline
    \end{tabular}
  \end{center}
\end{table}
\section{Concluding Remarks}\label{sec-7}

In this paper, we have proposed an data-adjusted joint significance mediation effects test procedure. The explicit expressions of size and power were derived. We also
have extended the AJS for performing small-scale multiple testing with FWER control. An adjusted Sobel-type confidence interval was presented. Some simulations and three real-world examples were used to illustrate the usefulness of our method. The method we propose provides a publicly accessible and user-friendly R package, called $\tt AdjMed$, which can be found at \url{ https://github.com/zhxmath/AdjMed}. The focus of this study was limited to a single mediator or a small number of mediators, and the inclusion of high-dimensional mediators was beyond the scope of this research.

There exist two possible directions for applying the proposed AJS test method in our future research.  (i) {\it Microbiome Mediation Analysis}.  Recently, increasing studies have studied the biological mechanisms
whether the microbiome play a mediating role between an  exposure
and a clinical outcome (\citeauthor{SohnandLi2019}, \citeyear{SohnandLi2019}). For improving the powers of mediation effect testing, it is desirable to use the  AJS test method when performing microbiome mediation analysis.
(ii) {\it Multiple-Mediator Testing with FDR control}. We have studied the theoretical and numerical performances of AJS test method for multiple testing with FWER control. It is useful to investigate the  AJS test for multiple mediators with FDR control in some applications.

\section*{Supplementary Material}
The Supplementary Material includes Figures S.1-S.3, Tables S.1-S.6, the proofs of Theorems \ref{Th1}-\ref{Th3}, the size of ASobel test, together with the manual for R package {\tt AdjMed}.
\section*{Acknowledgements}
We would like to thank the Editor, the Associate Editor and two reviewers for their
constructive and insightful comments that greatly improved the manuscript.

\bibliographystyle{natbib}
\bibliography{reference}
\newpage
\begin{center}
  Supplementary Materials for\\
  \large\bf Efficient Adjusted Joint Significance
Test and  Sobel-Type Confidence Interval for Mediation Effect
\end{center}

\vspace*{0.2in}



\appendix

\renewcommand{\theequation}{S.\arabic{equation}}
\renewcommand{\thefigure}{S.\arabic{figure}}
\renewcommand{\thelemma}{S.\arabic{lemma}}
\renewcommand{\thetable}{S.\arabic{table}}

\section{Proofs}
\setcounter{table}{0}
\setcounter{figure}{0}
In this section, we give the proof details of Theorems 1, 2,
and 3. \\

\noindent{\bf Proof of Theorem 1}. Under $H_{00}$, the size of the AJS test  is determined by
\begin{eqnarray}\label{A.10}
Size(AJS | H_{00}) &=& \mathbb{P}(P_{AJS} < \delta | H_{00})\\
&=& \mathbb{P}(P_{AJS} < \delta, T_{max} \geq \lambda_n | H_{00}) + \mathbb{P}(P_{AJS} < \delta, T_{max} > \lambda_n | H_{00}),\nonumber
\end{eqnarray}
where $T_{max} = \max\{|T_\alpha|, |T_\beta|\}$. 
It follows from  $\lim_{n\rightarrow \infty}\mathbb{P}(T_{max} \geq \lambda_n | H_{00}) = 0$  that
\begin{eqnarray}\label{A.11}
\lim_{n\rightarrow \infty}\mathbb{P}(P_{AJS} < \delta, T_{max} \geq \lambda_n| H_{00})
 &\leq& \lim_{n\rightarrow \infty}\mathbb{P}(T_{max} \geq \lambda_n|H_{00})\\
 &=&0.\nonumber
\end{eqnarray}
The distribution of $P_{JS}^2$ is noted to be a uniform distribution on the interval $(0, 1)$, {with probability approaching one}, under $H_{00}$ we can deduce that 
\begin{eqnarray}\label{A.12}
\lim_{n\rightarrow \infty}\mathbb{P}(P_{AJS} < \delta, T_{max} < \lambda_n|H_{00}) &=& \lim_{n\rightarrow \infty}\mathbb{P}(P_{JS}^2 < \delta, T_{max} < \lambda_n |H_{00})\\
&=&\lim_{n\rightarrow \infty}\mathbb{P}(P_{JS}^2 < \delta|H_{00})\nonumber\\
&=&\delta,\nonumber
\end{eqnarray}
{where the second equality holds because of the $\lim_{n\rightarrow \infty}\mathbb{P}\big(T_{max} < \lambda_n |H_{00}\big) = 1$}.
 The equations (\ref{A.10}), (\ref{A.11}) and (\ref{A.12}) imply that the asymptotic  size of AJS given $H_{00}$ is bounded by $\delta$.

Under $H_{10}$, the size of the AJS test is denoted as
\begin{eqnarray}\label{A.13}
Size(AJS | H_{10}) &=& \mathbb{P}(P_{AJS} < \delta | H_{10})\\
&=& \mathbb{P}(P_{AJS} < \delta, T_{max} \geq \lambda_n | H_{10}) + \mathbb{P}(P_{AJS} < \delta, T_{max} < \lambda_n | H_{10}).\nonumber
\end{eqnarray}
Under $H_{10}$, {with probability approaching one}, we can derive that
\begin{eqnarray}\label{A.14}
\lim_{n\rightarrow \infty}\mathbb{P}(P_{AJS} < \delta, T_{max} \geq\lambda_n | H_{10}) &=& \lim_{n\rightarrow \infty}\mathbb{P}(P_{JS} < \delta, T_{max} \geq\lambda_n |H_{10})\\
&=&\lim_{n\rightarrow \infty}\mathbb{P}(P_{\beta} < \delta, T_{max} \geq\lambda_n|H_{10})\nonumber\\
&=&\lim_{n\rightarrow \infty}\mathbb{P}(P_{\beta} < \delta|H_{10})\nonumber\\
&=&\delta,\nonumber
\end{eqnarray}
and
\begin{eqnarray}\label{A.15}
\lim_{n\rightarrow \infty}\mathbb{P}(P_{AJS} < \delta, T_{max} < \lambda_n| H_{10}) &\leq& \lim_{n\rightarrow \infty}\mathbb{P}(T_{max} < \lambda_n| H_{10})\\
& =& 0,\nonumber
\end{eqnarray}
where the last equality is due to $\lim_{n\rightarrow \infty}\mathbb{P}\big(T_{max} < \lambda_n |H_{10}\big) = 0$.
In view of (\ref{A.13}), (\ref{A.14}) and (\ref{A.15}), {with probability approaching one}, we have
\begin{eqnarray*}
\lim\limits_{n\rightarrow \infty}{ Size(AJS|H_{10})} &=& \delta.
\end{eqnarray*}
The size of the AJS test under $H_{01}$ is bounded by $\delta$ in a similar manner. This ends the proof.

\vspace{1cm}
 \noindent{\bf Proof of Theorem 2}. For the multiple testing problem, we deduce the FWER of our proposed AJS method under the significance level $\delta$. By the definition of FWER, we can derive  that
\begin{eqnarray*}
&&FWER(AJS)= \mathbb{P}\left(\bigcup_{k=1}^d\left\{P_{AJS}^{(k)} < \delta/d\right\}~\vline~ H_{0}^{(k)}, k=1,\cdots,d\right)\\
&&\leq \sum_{k=1}^d \mathbb{P}\left(P_{AJS}^{(k)} < \delta/d~\vline~ H_{0}^{(k)}\right)\\
&&=\sum_{k\in \Omega_{00}} \mathbb{P}\left(P_{AJS}^{(k)} < \delta/d~\vline~ H_{00}^{(k)}\right) + \sum_{k\in \Omega_{10}}\mathbb{P}\left(P_{AJS}^{(k)} < \delta/d~\vline~ H_{10}^{(k)}\right) + \sum_{k\in \Omega_{01}} \mathbb{P}\left(P_{AJS}^{(k)} < \delta/d~\vline~ H_{01}^{(k)}\right),
\end{eqnarray*}
where $\Omega_{00}= \{k: H^{(k)}_{00}, k=1,\cdots,d\}$, $\Omega_{10}= \{k: H^{(k)}_{10}, k=1,\cdots,d\}$ and $\Omega_{01}= \{k: H^{(k)}_{01}, k=1,\cdots,d\}$.
In view of the fact that
\begin{eqnarray}\label{2R-19}
\sum_{k\in \Omega_{00}} \mathbb{P}\left(P_{AJS}^{(k)} < \delta/d~\vline~ H_{00}^{(k)}\right) = R_1 + R_2,
\end{eqnarray}
where
\begin{eqnarray*}
{R_1} &=&{\sum_{k\in \Omega_{00}}\mathbb{P}\left(P_{AJS}^{(k)} < \delta/d, T_{max,k} \geq \lambda_n ~\vline~ H_{00}^{(k)}\right)},\\
{R_2} &=&{\sum_{k\in \Omega_{00}}\mathbb{P}\left(P_{AJS}^{(k)} < \delta/d, T_{max,k} < \lambda_n ~\vline~ H_{00}^{(k)}\right)},
\end{eqnarray*}
where $T_{max,k} = \max(|T_{\alpha_k}|, |T_{\beta_k}|)$.  In view of the fact that $\lim_{n\rightarrow \infty} \mathbb{P}(T_{max,k} \geq \lambda_n ~\vline~ H_{00}^{(k)})=0$,
the term $R_1$ satisfies the following expression:
\begin{eqnarray}\label{R1-20}
\lim_{n\rightarrow \infty}R_1
&\leq& \sum_{k\in \Omega_{00}} \lim_{n\rightarrow \infty} \mathbb{P}\left(T_{max,k} \geq \lambda_n ~\vline~ H_{00}^{(k)}\right)\\
&=&0.\nonumber
\end{eqnarray}
In addition, {with probability approaching one}, we get
\begin{eqnarray}\label{R1-21}
\lim_{n\rightarrow \infty}R_2&=& \sum_{k\in \Omega_{00}} \lim_{n\rightarrow \infty}\mathbb{P}\left(P_{AJS}^{(k)} < \delta/d,T_{max,k} < \lambda_n ~\vline~   H_{00}^{(k)}\right)\nonumber\\
&=&\sum_{k\in \Omega_{00}} \lim_{n\rightarrow \infty}\mathbb{P}\left(\{P_{JS}^{(k)}\}^2 < \delta/d, T_{max,k} < \lambda_n ~\vline~  H_{00}^{(k)}\right)\nonumber\\
&=&\sum_{k\in \Omega_{00}} \lim_{n\rightarrow \infty}\mathbb{P}\left(\{P_{JS}^{(k)}\}^2 < \delta/d ~\vline~  H_{00}^{(k)}\right)\nonumber\\
&=& |\Omega_{00}|\frac{\delta}{d}.
\end{eqnarray}
Hence, {with probability approaching one}, it follows from (\ref{2R-19}), (\ref{R1-20}) and (\ref{R1-21}) that
\begin{eqnarray}
\lim_{n\rightarrow \infty} \sum_{k\in \Omega_{00}} \mathbb{P}\left(P_{AJS}^{(k)} < \delta/d~\vline~ H_{00}^{(k)}\right) &\leq& |\Omega_{00}| \frac{\delta}{d}.
\end{eqnarray}

Next, we focus on the asymptotic upper bound of $\sum_{k=1}^d \mathbb{P}\left(P_{AJS}^{(k)} < \delta/d~\vline~ H_{10}^{(k)}\right)$ to control the FWER. To be specific, we note that
\begin{eqnarray}\label{AQ-23}
\sum_{k\in \Omega_{10}} \mathbb{P}\left(P_{AJS}^{(k)} < \delta/d~\vline~ H_{10}^{(k)}\right) = R_3 + R_4,
\end{eqnarray}
where
\begin{eqnarray*}
{R_3} &=&{\sum_{k\in \Omega_{10}}\mathbb{P}\left(P_{AJS}^{(k)} < \delta/d, T_{max,k} \geq \lambda_n ~\vline~ H_{10}^{(k)}\right)},\\
{R_4} &=&{\sum_{k\in \Omega_{10}}\mathbb{P}\left(P_{AJS}^{(k)} < \delta/d, T_{max,k} < \lambda_n ~\vline~ H_{10}^{(k)}\right)}.
\end{eqnarray*}
{ With probability approaching one,} some direct deductions lead to that
\begin{eqnarray}\label{AQ-24}
\lim_{n\rightarrow \infty}R_3&=& \lim_{n\rightarrow \infty}\sum_{k\in \Omega_{10}}\mathbb{P}\left(P_{AJS}^{(k)} < \delta/d, T_{max,k} \geq \lambda_n ~\vline~ H_{10}^{(k)}\right)\nonumber\\
&=& \lim_{n\rightarrow \infty}\sum_{k\in \Omega_{10}} \mathbb{P}\left( P_{JS}^{(k)} < \delta/d, T_{max,k} \geq \lambda_n  ~\vline~   H_{10}^{(k)}\right)\nonumber\\
&\leq& \lim_{n\rightarrow \infty}\sum_{k\in \Omega_{10}} \mathbb{P}\left( P_{JS}^{(k)} < \delta/d ~\vline~   H_{10}^{(k)}\right)\nonumber\\
&=&\frac{\delta}{d} |\Omega_{10}|,
\end{eqnarray}
and
\begin{eqnarray}\label{AQ-25}
\lim_{n\rightarrow \infty}R_4&=& \lim_{n\rightarrow \infty}{\sum_{k\in \Omega_{10}}\mathbb{P}\left(P_{AJS}^{(k)} < \delta/d, T_{max,k} < \lambda_n ~\vline~ H_{10}^{(k)}\right)}\\
&\leq&\lim_{n\rightarrow \infty}\sum_{k\in \Omega_{10}} \mathbb{P}\left(T_{max,k} < \lambda_n ~\vline~ H_{10}^{(k)}\right)\nonumber\\
&=&0,\nonumber
\end{eqnarray}
where the last equality is from $\lim_{n\rightarrow \infty} \mathbb{P} (T_{max,k} < \lambda_n | H_{10}^{(k)})=0$. 
Based on (\ref{AQ-23}), (\ref{AQ-24}) and (\ref{AQ-25}), we have
\begin{eqnarray*}
\lim_{n\rightarrow \infty}\sum_{k\in \Omega_{10}} \mathbb{P}\left(P_{AJS}^{(k)} < \delta/d~\vline~ H_{10}^{(k)}\right)
&\leq& |\Omega_{10}|\frac{\delta}{d}.
\end{eqnarray*}
In a similar procedure, {with probability approaching one,} it can be demonstrated that
$\lim_{n\rightarrow \infty}\sum_{k\in \Omega_{01}} \mathbb{P}(P_{AJS}^{(k)} < \delta/d~|~ H_{01}^{(k)}) \leq |\Omega_{01}| {\delta}/{d}$. Therefore, {with probability approaching one},  the asymptotic upper bound of AJS's FWER is
\begin{eqnarray*}
\lim_{n\rightarrow \infty}{\rm FWER(AJS)} &\leq& |\Omega_{00}| \frac{\delta}{d} + |\Omega_{10}| \frac{\delta}{d} + |\Omega_{01}| \frac{\delta}{d}\\
&=&\delta,
\end{eqnarray*}
where the last equality is due to $|\Omega_{00}| + |\Omega_{10}| + |\Omega_{01}| = d$. This completes the proof.

\vspace{1.2cm}

\noindent{\bf Proof of Theorem 3}.  By the definition of  ASobel-type confidence interval, we have
\begin{eqnarray}\label{S-17}
&&\mathbb{P}(\alpha\beta \in {\rm CI_{ASobel}}|\alpha=\beta=0) \\
&&= \mathbb{P}(\alpha\beta \in {\rm CI_{ASobel}}, T_{max} \geq \lambda_n |\alpha=\beta=0) + \mathbb{P}(\alpha\beta \in {\rm CI_{ASobel}}, T_{max} < \lambda_n |\alpha=\beta=0),\nonumber
\end{eqnarray}
where the term $T_{max} = \max(|T_\alpha|, |T_\beta|) $ satisfying $\lim_{n\rightarrow \infty}\mathbb{P}(T_{max} \geq \lambda_n | \alpha=\beta=0) = 0$ and $\lim_{n\rightarrow \infty}\mathbb{P}(T_{max} <\lambda_n | \alpha=\beta=0) = 1$. From the definition of ${\rm CI_{ASobel}}$, {with probability approaching one}, it is straightforward to derive the following expressions:
\begin{eqnarray}\label{S-18}
\lim_{n\rightarrow \infty}\mathbb{P}\left(\alpha\beta \in {\rm CI_{ASobel}}, T_{max} \geq \lambda_n | \alpha=\beta=0\right) 
&\leq& \lim_{n\rightarrow \infty} \mathbb{P}\left(T_{max} \geq \lambda_n | \alpha=\beta=0\right) \nonumber\\
&=&0,
\end{eqnarray}
and
\begin{eqnarray}\label{S-19}
\lim_{n\rightarrow \infty}\mathbb{P}(\alpha\beta \in {\rm CI_{ASobel}}, T_{max} < \lambda_n| \alpha=\beta=0)  
&=& 1-\lim_{n\rightarrow \infty}\mathbb{P}(|T_{Sobel}| > N_{1-\delta/2}(0,1/4)|\alpha=\beta=0)\nonumber\\
&=&1-\delta,
\end{eqnarray}
where $T_{Sobel} = (\hat{\alpha}\hat{\beta}-\alpha\beta)/\hat{\sigma}_{\alpha\beta}$, and
$N_{1-\delta/2}(0,1/4)$ denotes the $(1-\delta/2)$-quantile of $N(0,1/4)$.
From (\ref{S-17}), (\ref{S-18}) and (\ref{S-19}), {with probability approaching one}, we can derive  that
\begin{eqnarray*}
\lim_{n\rightarrow \infty}\mathbb{P}(\alpha\beta \in {\rm CI_{ASobel}}|\alpha=\beta=0) = 1-\delta.
\end{eqnarray*}

Next, we derive the asymptotic coverage probability of  ${\rm CI_{ASobel}}$ under $\alpha=0,\beta\neq 0$. Note that 
\begin{eqnarray}\label{S-20}
&&\mathbb{P}(\alpha\beta \in {\rm CI_{ASobel}}|\alpha=0,\beta\neq 0) \\
&&= \mathbb{P}(\alpha\beta \in {\rm CI_{ASobel}}, T_{max} \geq \lambda_n |\alpha=0,\beta\neq 0) + \mathbb{P}(\alpha\beta \in {\rm CI_{ASobel}}, T_{max} < \lambda_n |\alpha=0,\beta\neq 0).\nonumber
\end{eqnarray}
{With probability approaching one}, some calculations lead to that
\begin{eqnarray}\label{A.s5}
\lim_{n\rightarrow \infty}\mathbb{P}(\alpha\beta \in {\rm CI_{ASobel}}, T_{max} \geq \lambda_n | \alpha=0,\beta\neq 0) 
&=& 1-\lim_{n\rightarrow \infty}\mathbb{P}(|T_{Sobel}| > N_{1-\delta/2}(0,1)|\alpha=0,\beta\neq 0)\nonumber\\
&=&1-\delta,
\end{eqnarray}
together with
\begin{eqnarray}\label{A.s6}
\lim_{n\rightarrow \infty}\mathbb{P}(\alpha\beta \in {\rm CI_{ASobel}}, T_{max} < \lambda_n | \alpha=0,\beta\neq 0) 
&\leq& \lim_{n\rightarrow \infty}\mathbb{P}(T_{max} < \lambda_n | \alpha=0,\beta\neq 0)\nonumber\\
&=&0,
\end{eqnarray}
where the last equality is derived from
\begin{eqnarray*}
\lim_{n\rightarrow \infty}\mathbb{P}(T_{max} \geq \lambda_n | \alpha=0,\beta\neq 0)=1-\lim_{n\rightarrow \infty}\mathbb{P}(T_{max} < \lambda_n | \alpha=0,\beta\neq 0)=1.
\end{eqnarray*}
In view of (\ref{S-20}), (\ref{A.s5}) and (\ref{A.s6}), {with probability approaching one}, under $\alpha=0,\beta\neq 0$ we can derive the   asymptotic coverage probability of  ${\rm CI_{ASobel}}$ as
\begin{eqnarray*}
\lim_{n\rightarrow \infty}\mathbb{P}(\alpha\beta \in {\rm CI_{ASobel}}|\alpha=0,\beta\neq 0)
&=&1-\delta.
\end{eqnarray*}
In a similar way, {with probability approaching one}, it can be deduced that 
$$\lim_{n\rightarrow \infty}\mathbb{P}(\alpha\beta \in {\rm CI_{ASobel}}|\alpha\neq0,\beta=0)
= 1-\delta.$$

Lastly, we focus on deducing the asymptotic coverage probability of  ${\rm CI_{ASobel}}$ under $\alpha\neq0,\beta\neq 0$. Specifically, we can derive that
\begin{eqnarray}\label{S-23}
&&\mathbb{P}(\alpha\beta \in {\rm CI_{ASobel}}|\alpha\neq0,\beta\neq 0) \\
&&= \mathbb{P}(\alpha\beta \in {\rm CI_{ASobel}}, T_{max} \geq \lambda_n |\alpha\neq0,\beta\neq 0) + \mathbb{P}(\alpha\beta \in {\rm CI_{ASobel}}, T_{max} < \lambda_n |\alpha\neq0,\beta\neq 0),\nonumber
\end{eqnarray}
where 
\begin{eqnarray}\label{S-24}
\lim_{n\rightarrow \infty}\mathbb{P}(\alpha\beta \in {\rm CI_{ASobel}}, T_{max} \geq \lambda_n | \alpha\neq0,\beta\neq 0) 
&=& 1-\lim_{n\rightarrow \infty}\mathbb{P}(|T_{Sobel}| > N_{1-\delta/2}(0,1)|\alpha\neq0,\beta\neq 0)\nonumber\\
&=&1-\delta,
\end{eqnarray}
and
\begin{eqnarray}\label{S-25}
\lim_{n\rightarrow \infty}\mathbb{P}(\alpha\beta \in {\rm CI_{ASobel}}, T_{max} < \lambda_n | \alpha\neq0,\beta\neq 0) 
&\leq&\lim_{n\rightarrow \infty}\mathbb{P}(T_{max} < \lambda_n | \alpha\neq0,\beta\neq 0)\nonumber\\
&=&0.
\end{eqnarray}
Here the last equality is due to
\begin{eqnarray*}
\lim_{n\rightarrow \infty}\mathbb{P}(T_{max} \geq \lambda_n | \alpha\neq0,\beta\neq 0)=1-\lim_{n\rightarrow \infty}\mathbb{P}(T_{max} < \lambda_n | \alpha\neq0,\beta\neq 0)=1.
\end{eqnarray*}
{With probability approaching one}, it follows from (\ref{S-23}), (\ref{S-24}) and (\ref{S-25}) that  $\lim_{n\rightarrow \infty}\mathbb{P}(\alpha\beta \in {\rm CI_{ASobel}}|\alpha\neq0,\beta\neq 0)
= 1-\delta.$ This ends the proof.

\section{The Size of ASobel Test}
In this section, we provide the proof details for the size  of ASobel test. 
Under $H_{00}$, it follows from the decision rule of the ASobel test that
\begin{eqnarray}\label{A.1}
Size(ASobel | H_{00}) &=& \mathbb{P}(P_{ASobel} < \delta | H_{00})\\
&=& \mathbb{P}(P_{ASobel} < \delta, T_{max} < \lambda_n| H_{00}) + \mathbb{P}(P_{ASobel} < \delta, T_{max} \geq \lambda_n | H_{00}),\nonumber
\end{eqnarray}
where $T_{max} = \max(|T_\alpha|, |T_\beta|) $.
From the definition of $P_{ASobel}$ and  $\lim_{n\rightarrow \infty}\mathbb{P}(T_{max} \geq \lambda_n | H_{00}) = 0$, {with probability approaching one}, it is straightforward to derive the following expressions:
\begin{eqnarray}\label{A.2}
\lim_{n\rightarrow \infty}\mathbb{P}(P_{ASobel} < \delta, T_{max} \geq \lambda_n| H_{00}) &\leq& \lim_{n\rightarrow \infty}\mathbb{P}(T_{max} \geq \lambda_n| H_{00})\nonumber\\
&=&0,
\end{eqnarray}
and
\begin{eqnarray}\label{A.3}
\lim_{n\rightarrow \infty}\mathbb{P}(P_{ASobel} < \delta, T_{max} < \lambda_n| H_{00}) &=& \lim_{n\rightarrow \infty}\mathbb{P}\Big( 2\{1-\Phi_{N(0,1/4)}(|T_{Sobel}|)\} < \delta ~\vline~ H_{00}\Big)\nonumber\\
&=& \lim_{n\rightarrow \infty}\mathbb{P}(|T_{Sobel}| > N_{1-\delta/2}(0,1/4)|H_{00})\nonumber\\
&=&\delta,
\end{eqnarray}
where $\Phi_{N(0,1/4)}(\cdot)$ is the cumulative distribution function of $N(0,1/4)$, and
$N_{1-\delta/2}(0,1/4)$ denotes the $(1-\delta/2)$-quantile of $N(0,1/4)$.
From (\ref{A.1}), (\ref{A.2}) and (\ref{A.3}), {with probability approaching one}, we can derive  that
\begin{eqnarray*}
\lim_{n\rightarrow \infty}Size(ASobel | H_{00}) = \delta.
\end{eqnarray*}

 Under $H_{10}$, we have
\begin{eqnarray}\label{A.4}
Size(ASobel | H_{10}) &=& \mathbb{P}(P_{ASobel} < \delta | H_{10})\\
&=& \mathbb{P}(P_{ASobel} < \delta, T_{max} \geq \lambda_n | H_{10}) + \mathbb{P}(P_{ASobel} < \delta, T_{max} < \lambda_n | H_{10}).\nonumber
\end{eqnarray}
{With probability approaching one}, some calculations lead to that
\begin{eqnarray}\label{A.5}
\lim_{n\rightarrow \infty}\mathbb{P}(P_{ASobel} < \delta, T_{max} \geq \lambda_n| H_{10}) &=& \lim_{n\rightarrow \infty}\mathbb{P}\Big( 2\{1-\Phi_{N(0,1)}(|T_{Sobel}|)\} < \delta |H_{10}\Big)\nonumber\\
&=& \lim_{n\rightarrow \infty} \mathbb{P}(|T_{Sobel}| > N_{1-\delta/2}(0,1)|H_{10})\nonumber\\
&=&\delta,
\end{eqnarray}
together with
\begin{eqnarray}\label{A.6}
\lim_{n\rightarrow \infty}\mathbb{P}(P_{ASobel} < \delta, T_{max} < \lambda_n | H_{10}) &\leq& \lim_{n\rightarrow \infty}\mathbb{P}(T_{max} < \lambda_n | H_{10})\nonumber\\
&=&0,
\end{eqnarray}
where
$N_{1-\delta/2}(0,1)$ is the $(1-\delta/2)$-quantile of $N(0,1)$. In view of (\ref{A.4}), (\ref{A.5}) and (\ref{A.6}), {with probability approaching one}, we get 
\begin{eqnarray*}
\lim_{n\rightarrow \infty}Size(ASobel | H_{10})
&=&\delta.
\end{eqnarray*}
 In a similar way,  {with probability approaching one, we can conclude that the size of ASobel given $H_{01}$ is  $\delta$ as $n$ tends to infinity}. The proof for the size of ASobel test ends here.

In what follows, we provide more insights about the size  of Sobel test. Note that the size of traditional Sobel test under $H_{00}$ is
\begin{eqnarray*}
Size(Sobel | H_{00}) &=& \mathbb{P}\left(2\{1-\Phi_{N(0,1)}(|T_{Sobel}|) < \delta | H_{00}\right)\nonumber\\
&=&\mathbb{P}\left(|T_{Sobel}| >  N_{1-\delta/2}(0,1)| H_{00}\right)\nonumber\\
&=& 2 \{1- \Phi_{N(0,1/4)}(N_{1-\delta/2}(0,1))\}.
\end{eqnarray*}
By deducing the difference between $Size(Sobel | H_{00})$ and the significance level $\delta$, we have
\begin{eqnarray*}
Size(Sobel | H_{00}) - \delta &=& 2 \{1- \Phi_{N(0,1/4)}(N_{1-\delta/2}(0,1))\} - \delta\\
&<&0,
\end{eqnarray*}
which provides a theoretical explanation about the conservative performance of  traditional Sobel test under $H_{00}$.   Therefore,  the proposed ASobel test has a significant improvement over  traditional Sobel test in terms of size under $H_{00}$. 

\section{Figures S.1-S.3 and Tables S.1-S.6}
The following section provides Figures S.1-S.3, together with Tables S.1-S.6 for the simulation  section discussed in the main paper.

\begin{figure}[htp] 
  \centering
  \begin{subfigure}{0.4\textwidth}
    \includegraphics[width=\textwidth]{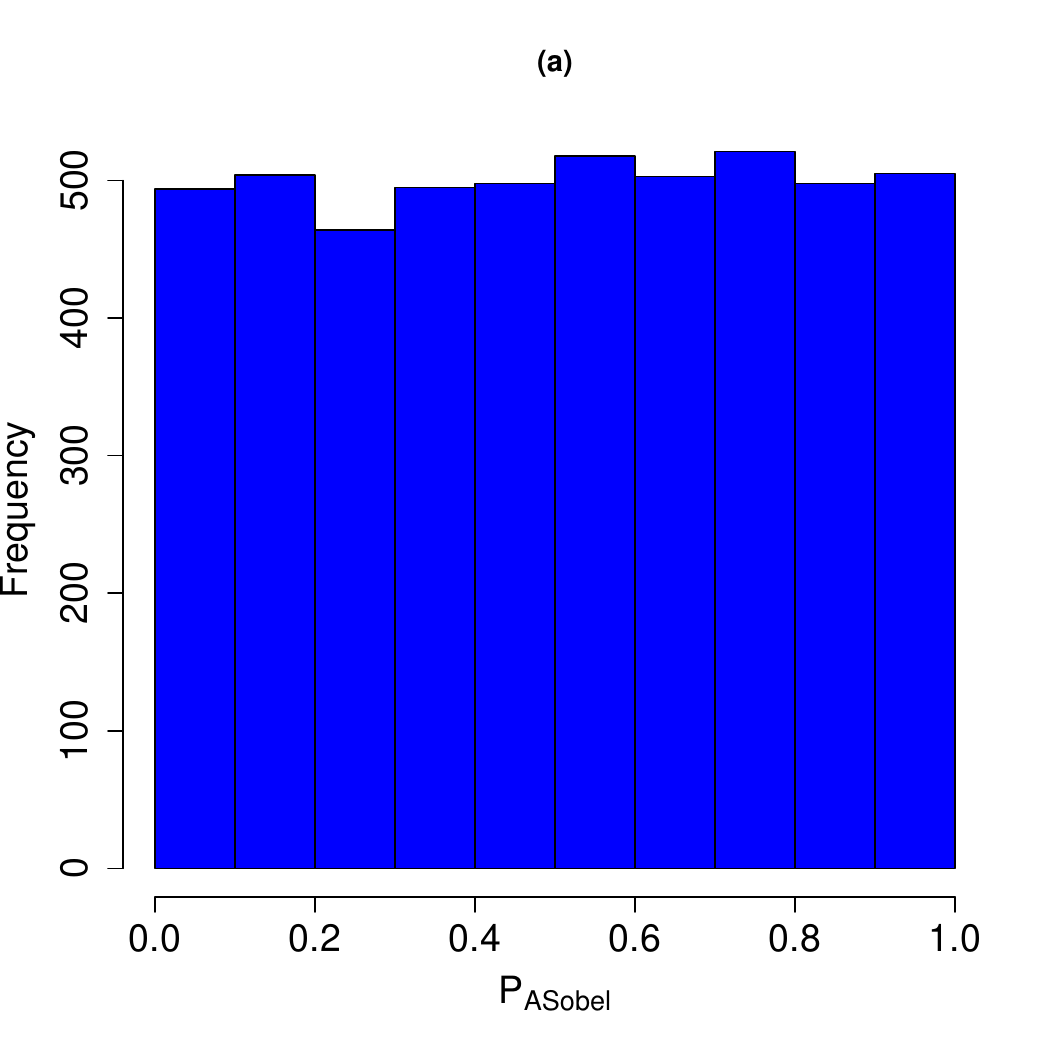}
    \caption{The p-values of ASobel method.}
  \end{subfigure}
  \begin{subfigure}{0.4\textwidth}
    \includegraphics[width=\textwidth]{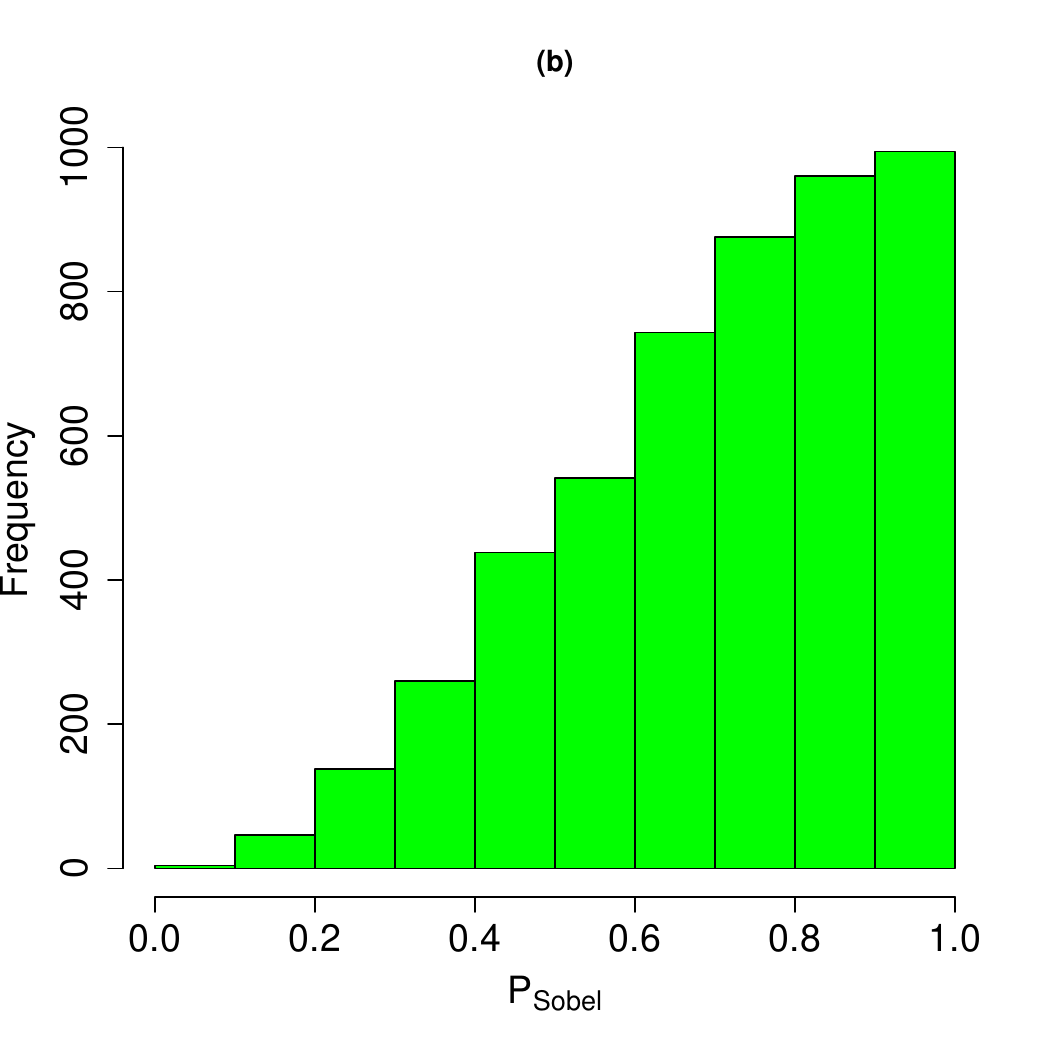}
    \caption{The p-values of Sobel method.}
  \end{subfigure}
 \vspace{-0.1cm}
\begin{center}
  \caption{The histogram of p-values for $(\alpha, \beta) = (0, 0)$ with $\lambda_n =  \sqrt{n}/\log(n)$.}
  \label{fig:4}
\end{center}
\end{figure}

\clearpage

\begin{table}[htp] 
  \begin{center}
    \caption{The size and power of hypothesis testing with logistic mediation model.}
    \label{tab:S1}
    \vspace{0.1in} \small
    \begin{tabular}{llcccccccccc}
      \hline
      & $(\alpha, \beta)$ & &Sobel &JS &Bootstrap&ASobel &AJS \\
      \hline
    $n=200$
  & (0, 0)           & &0         & 0.0030 & 0.0024 & 0.0486 & 0.0478\\
  & (0.5, 0)         & & 0.0368    & 0.0476 &0.0524 & 0.0368 & 0.0476\\
  & (0, 0.65)        & &0.0182    & 0.0486 &  0.0524 & 0.0506 &0.0724\\
  & (0.15, 0.20)     & &0.0350    &0.1354 &  0.0882 &0.2826& 0.3206\\
  & (0.25, 0.30)     & &0.2542    &0.4296 & 0.3600 &0.3680 &  0.4974\\ 
      \hline
    $n=500$
  & (0, 0)           & &0         & 0.0018 & 0.0020 & 0.0476 & 0.0484\\
  & (0.5, 0)         & &0.0448    & 0.0472 &0.0494  & 0.0448 & 0.0472\\
  & (0, 0.65)        & &0.0382    & 0.0530 & 0.0570 & 0.0398 & 0.0544\\
  & (0.15, 0.20)     & &0.2960    & 0.4840 & 0.3982 & 0.6492 & 0.6680\\
  & (0.25, 0.30)     & &0.7966    & 0.8344 & 0.8354 & 0.8046 & 0.8374\\ 
 \hline
    $n=1000$
  & (0, 0)           & &0         & 0.0018 & 0.0010 & 0.0518 & 0.0528\\
  & (0.5, 0)         & &0.0516    & 0.0532 & 0.0558 & 0.0516 & 0.0532\\
  & (0, 0.65)        & &0.0446    & 0.0510 &  0.0592 &0.0446 & 0.0510\\
  & (0.15, 0.20)     & &0.7384    & 0.8124 & 0.7854 & 0.8570 & 0.8708\\
  & (0.25, 0.30)     & &0.9866    & 0.9886 & 0.9886 &0.9866 & 0.9886\\ 
       \hline
    \end{tabular}
  \end{center}
\end{table}

\clearpage
\begin{table}[htp] 
  \begin{center}
    \caption{The size and power of hypothesis testing with Cox mediation model.}
    \label{tab:S2}
    \vspace{0.1in} \small
    \begin{tabular}{llcccccccccc}
      \hline
      & $(\alpha, \beta)$ & &Sobel &JS &Bootstrap&ASobel &AJS \\
      \hline
    $n=200$
  & (0, 0)      & &0         & 0.0030 & 0.0018 & 0.0428 & 0.0448\\
  & (0.5, 0)    & &0.0396    & 0.0498 & 0.0494 & 0.0396 & 0.0498\\
  & (0, 0.5)    & &0.0376    & 0.0510 & 0.0604 & 0.0382 &  0.0514\\
  & (0.15, 0.15)& & 0.0802    &0.2322 & 0.2024 & 0.3552 & 0.4302\\
  & (0.25, 0.25)& &0.5738    & 0.7494 & 0.7276 & 0.6522 & 0.7918\\ 
      \hline
    $n=500$
  & (0, 0)      & &0.0006    & 0.0044 & 0.0028 & 0.0546 & 0.0576\\
  & (0.5, 0)    & &0.0512    &0.0558 & 0.0576 & 0.0512 & 0.0558\\
  & (0, 0.5)    & &0.0482    & 0.0570 & 0.0572 & 0.0482 & 0.0570\\
  & (0.15, 0.15)& &0.5236    & 0.7266 & 0.7100 & 0.8156 & 0.8452\\
  & (0.25, 0.25)& &0.9920    & 0.9946 & 0.9946 & 0.9922 & 0.9948\\ 
 \hline
    $n=1000$
  & (0, 0)      & &0         & 0.0026 & 0.0016 & 0.0482 & 0.0480\\
  & (0.5, 0)    & &0.0484    & 0.0494 & 0.0520 & 0.0484 & 0.0494\\
  & (0, 0.5)    & &0.0432    & 0.0462 & 0.0498 & 0.0432 & 0.0462\\
  & (0.15, 0.15)& &0.9486    & 0.9754 & 0.9750 & 0.9786 & 0.9860\\
  & (0.25, 0.25)& &1    & 1 & 1 & 1 & 1\\ 
       \hline
    \end{tabular}
  \end{center}
\end{table}

\begin{figure}[htp] 
  \centering
  \begin{subfigure}{0.4\textwidth}
    \includegraphics[width=\textwidth]{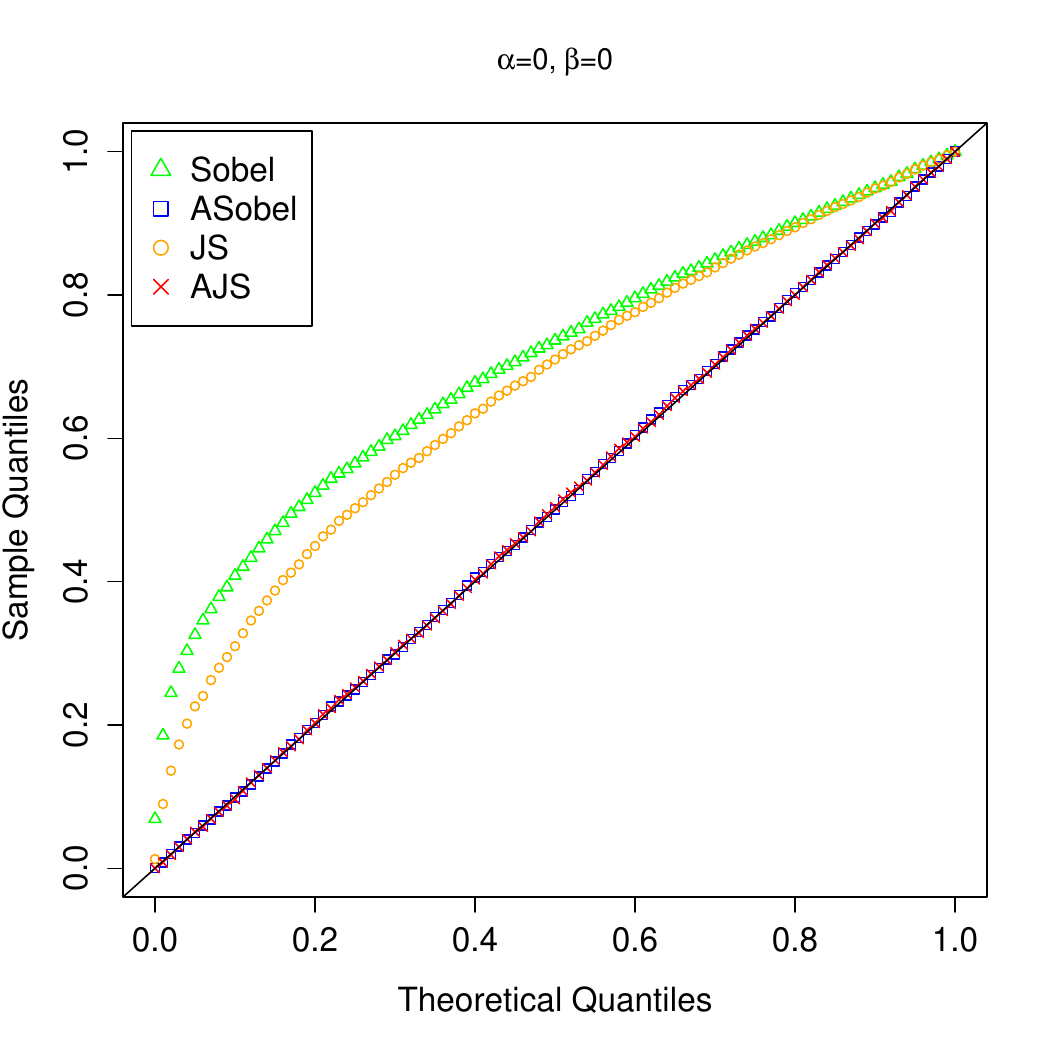}
    \caption{Q-Q plots of p-values under $H_{00}$.}
  \end{subfigure}
  \begin{subfigure}{0.4\textwidth}
    \includegraphics[width=\textwidth]{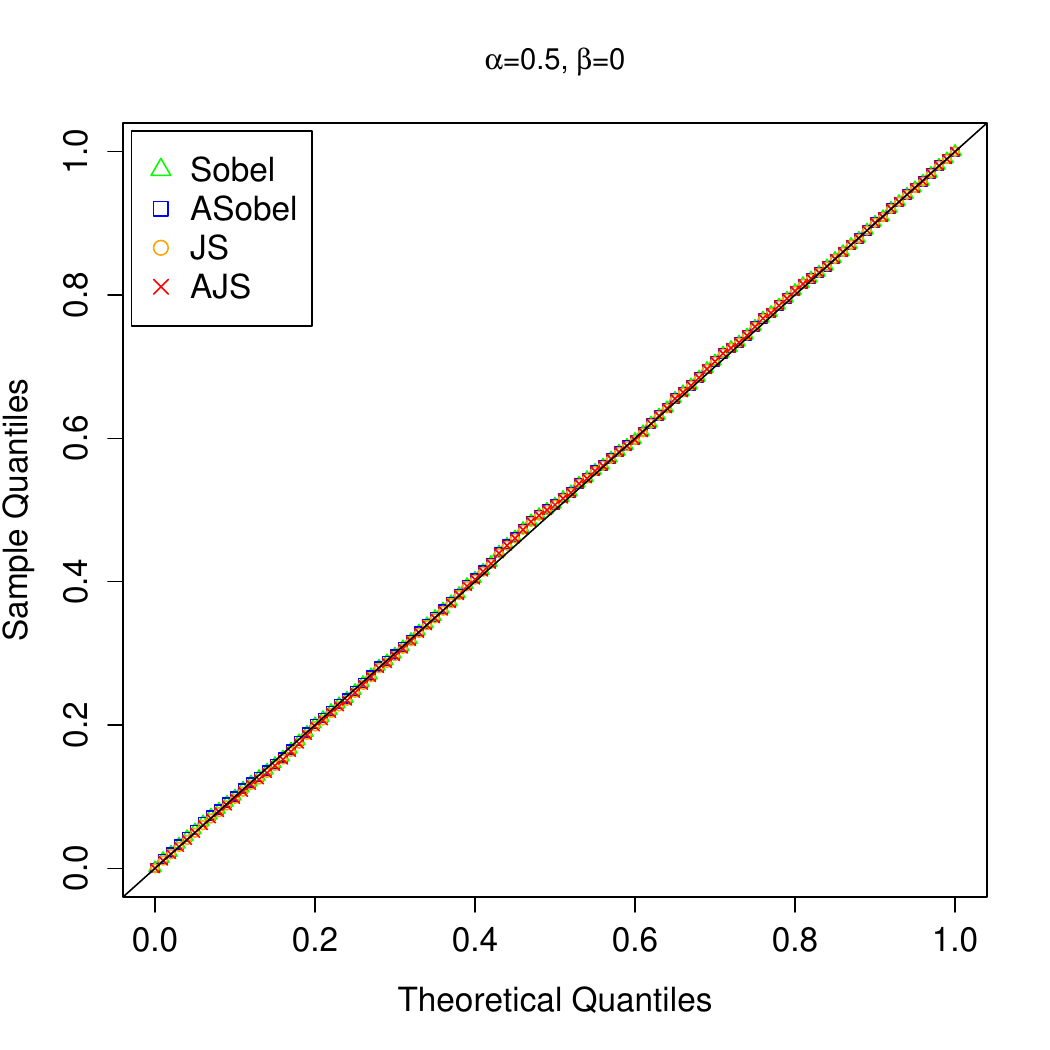}
    \caption{Q-Q plots of p-values under $H_{01}$.}
  \end{subfigure}
    \begin{subfigure}{0.4\textwidth}
    \includegraphics[width=\textwidth]{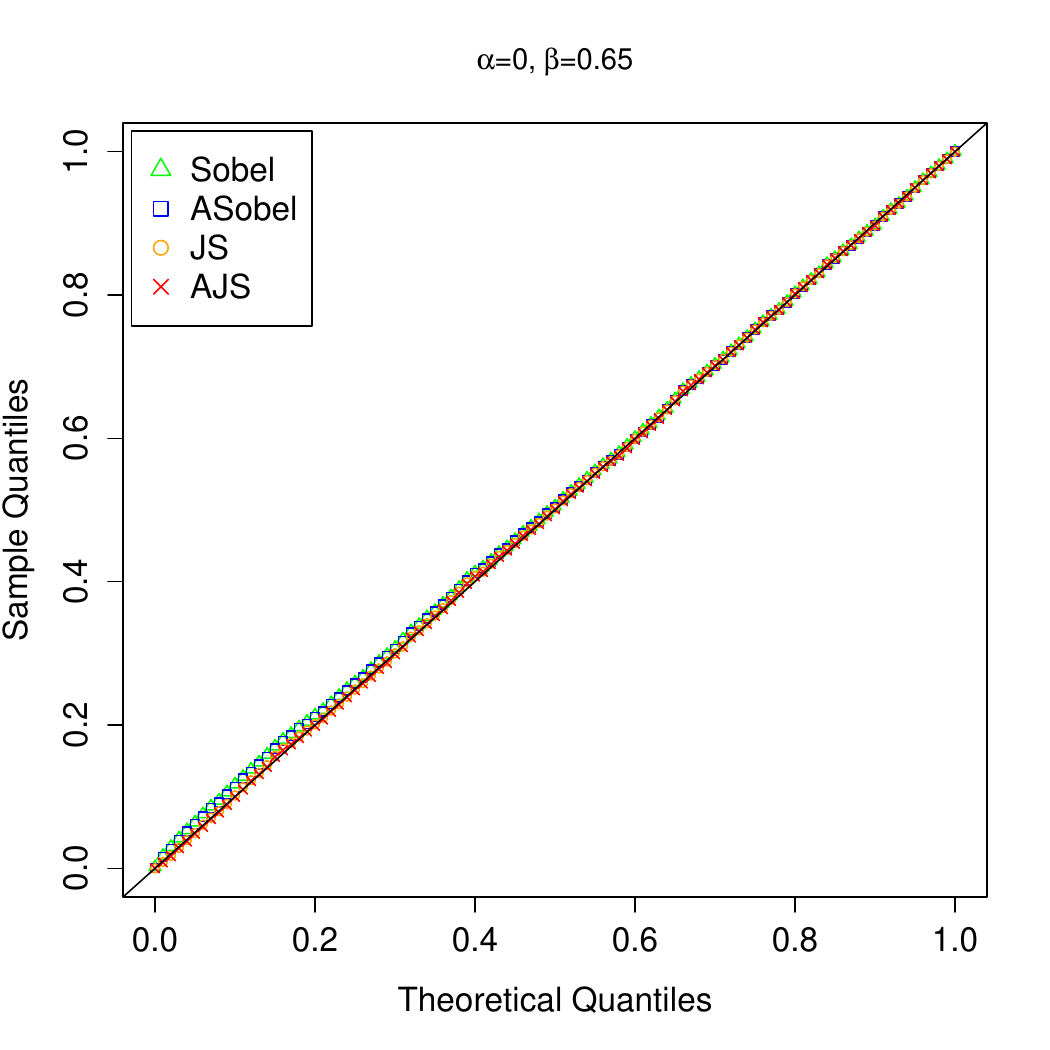}
    \caption{Q-Q plots of p-values under $H_{10}$.}
  \end{subfigure}
  \begin{subfigure}{0.4\textwidth}
    \includegraphics[width=\textwidth]{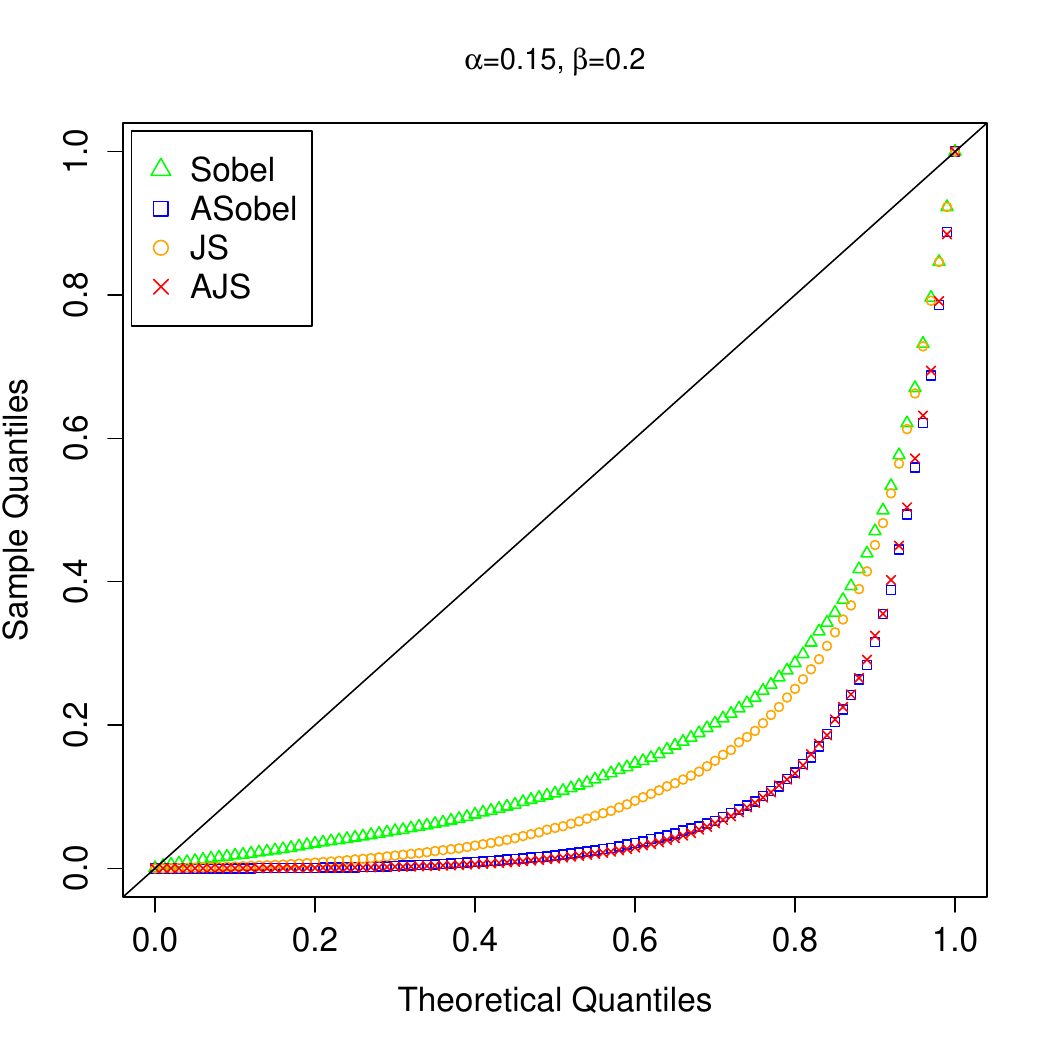}
    \caption{Q-Q plots of p-values under $H_{A}$.}
  \end{subfigure}
 \vspace{-0.1cm}
\begin{center}
  \caption{Q-Q plots of p-values under logistic mediation model with $n = 500$.}
  \label{fig:S1}
\end{center}
\end{figure}

\clearpage
\begin{figure}[htp] 
  \centering
  \begin{subfigure}{0.4\textwidth}
    \includegraphics[width=\textwidth]{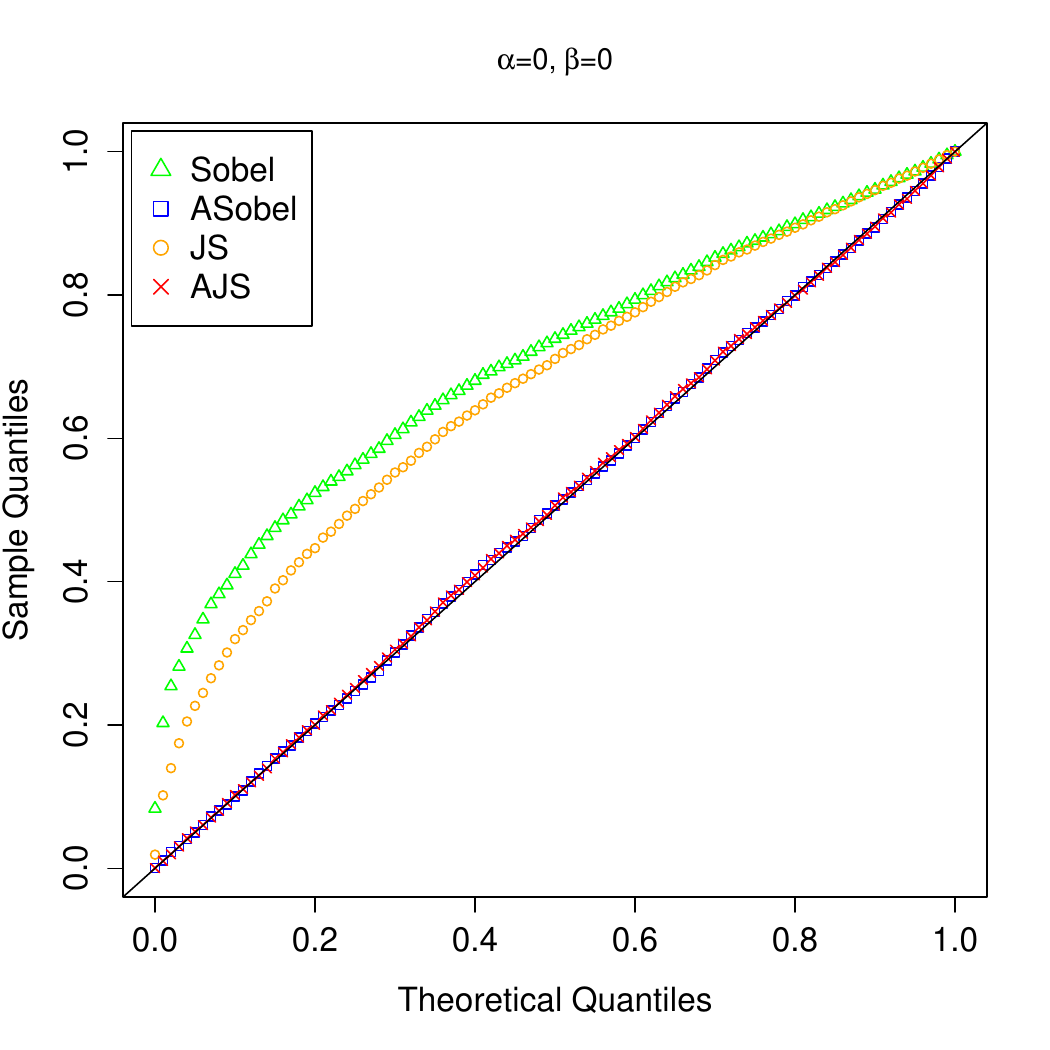}
    \caption{Q-Q plots of p-values under $H_{00}$.}
  \end{subfigure}
  \begin{subfigure}{0.4\textwidth}
    \includegraphics[width=\textwidth]{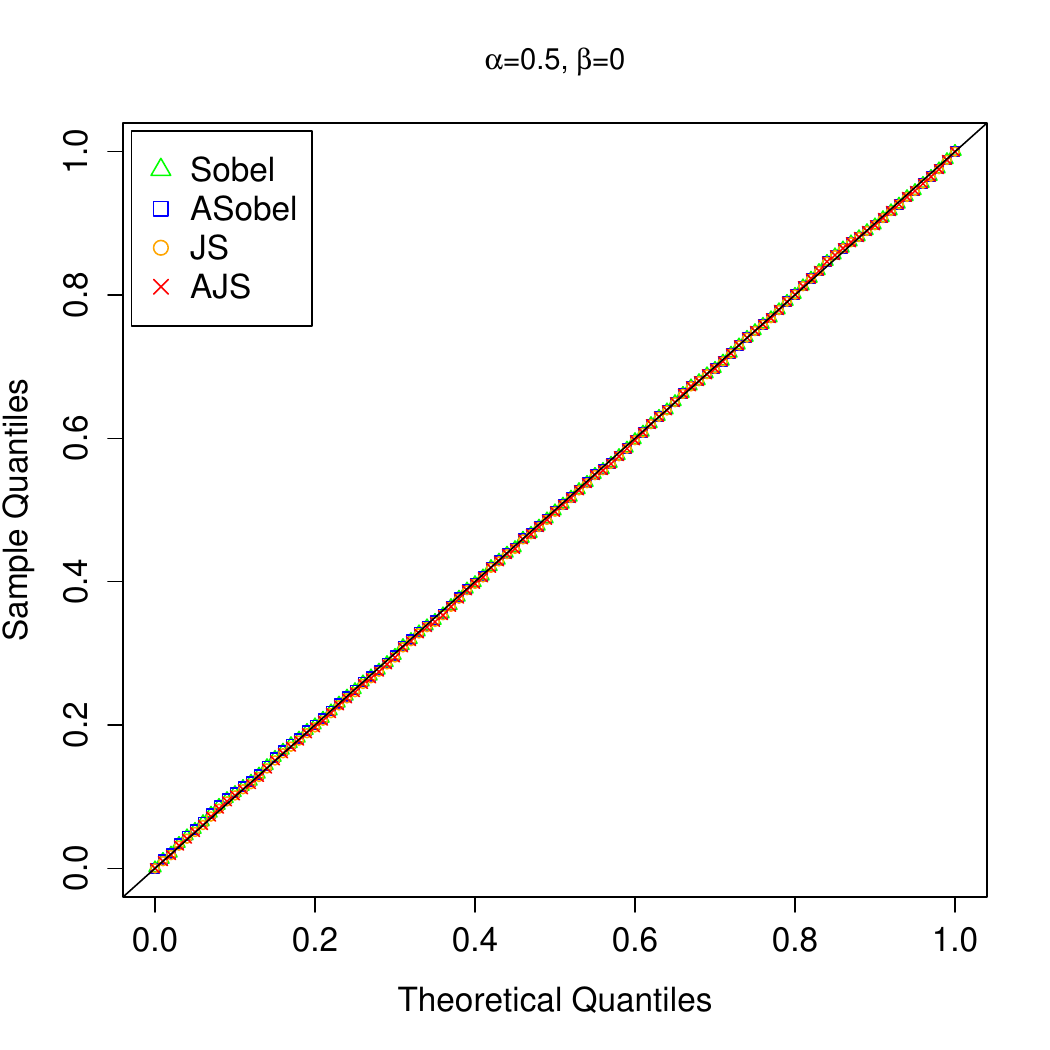}
    \caption{Q-Q plots of p-values under $H_{01}$.}
  \end{subfigure}
    \begin{subfigure}{0.4\textwidth}
    \includegraphics[width=\textwidth]{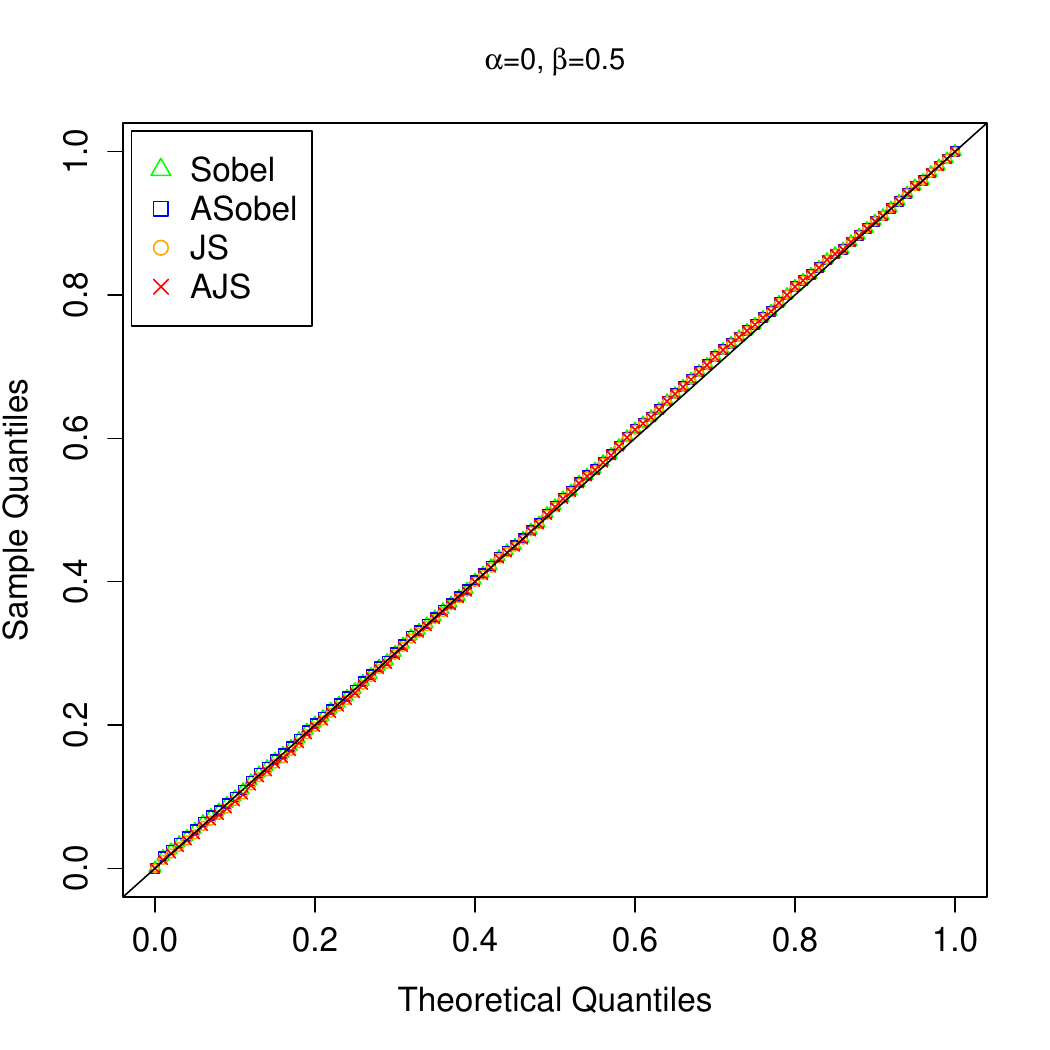}
    \caption{Q-Q plots of p-values under $H_{10}$.}
  \end{subfigure}
  \begin{subfigure}{0.4\textwidth}
    \includegraphics[width=\textwidth]{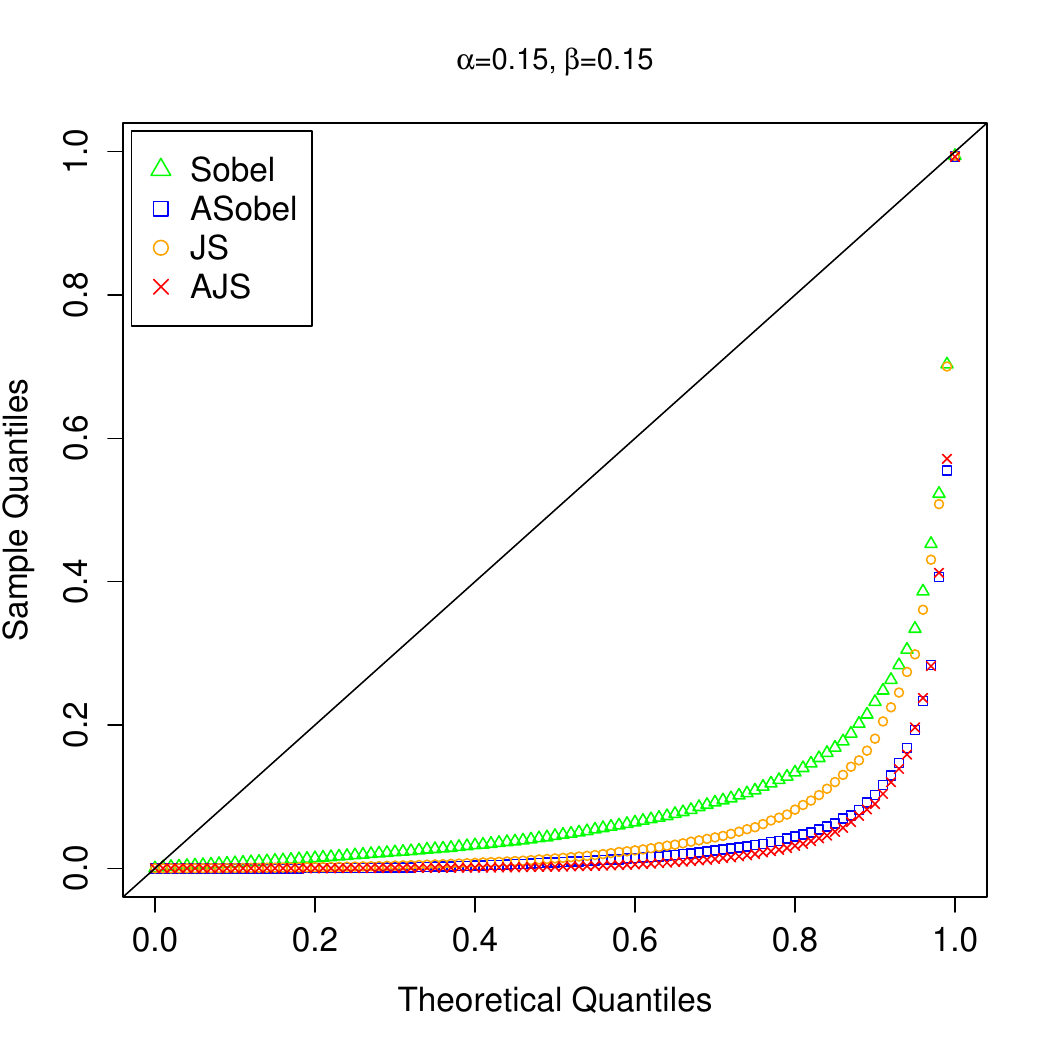}
    \caption{Q-Q plots of p-values under $H_{A}$.}
  \end{subfigure}
 \vspace{-0.1cm}
\begin{center}
  \caption{Q-Q plots of p-values under Cox mediation model with $n = 500$.}
  \label{fig:S1}
\end{center}
\end{figure}

\clearpage
\begin{table}[htp] 
  \begin{center}
    \caption{The FWER and power of multiple testing with logistic mediation model.}
    \label{tab:S3}
    \vspace{0.1in} \small
    \begin{tabular}{llcccccccccc}
      \hline
  & Dimension & &Sobel &JS & JT\_Comp & DACT &Bootstrap & AJS\\
      \hline
     $n=200$
   & $d=10$ &FWER &0.0030 &0.0058  & 0.1480 & 0       & 0.0092 & 0.0438\\
      &     &Power&0.0722 &0.1586  & 0.1752 & 0.0155  & 0.1645 & 0.2055\\
   & $d=15$ &FWER &0.0020 &0.0050  & 0.1568 & 0.0008  & 0.0048 & 0.0474\\
      &     &Power&0.0546 &0.1318  & 0.1721 & 0.0454  & 0.1204 & 0.1676\\
   & $d=20$ &FWER &0.0016 &0.0038  & 0.1718 & 0.0020  & 0.0022 & 0.0500\\
      &     &Power&0.0401 &0.1072  & 0.1738 & 0.0740  & 0.0756 & 0.1398\\
      \hline
     $n=500$
   & $d=10$ &FWER &0.0046 &0.0094  & 0.1424 & 0       & 0.0136 & 0.0594\\
      &     &Power&0.3617 &0.4860  & 0.2967 & 0.0138  & 0.5004 & 0.5486\\
   & $d=15$ &FWER &0.0040 &0.0094  & 0.1576 & 0       & 0.0112 & 0.0592\\
      &     &Power&0.3263 &0.4522  & 0.3012 & 0.0640  & 0.4653 & 0.5116\\
   & $d=20$ &FWER &0.0026 & 0.0052  & 0.1650 & 0.0004  & 0.0088 & 0.0650\\
      &     &Power&0.2998 &0.4316 &0.3156 & 0.1651  & 0.4432 & 0.4887\\
      \hline
     $n=1000$
   & $d=10$ &FWER &0.0058 &0.0110  & 0.1432 & 0       & 0.0154 & 0.0468\\
      &     &Power&0.6410 &0.7287  & 0.4536 & 0.0067  & 0.7417 & 0.7677\\
   & $d=15$ &FWER &0.0040 &0.0068  & 0.1410 & 0       & 0.0114 & 0.0438\\
      &     &Power&0.6018 &0.6985  & 0.4526 & 0.0771  & 0.7139 & 0.7382\\
   & $d=20$ &FWER &0.0040 &0.0050  & 0.1710 & 0.0002  & 0.0092 & 0.0526\\
      &     &Power&0.5778 &0.6809  & 0.4772 & 0.2488  &  0.6993 & 0.7216\\
      \hline
    \end{tabular}
  \end{center}
\end{table}

\clearpage
\begin{table}[htp] 
  \begin{center}
    \caption{The FWER and power of multiple testing with Cox mediation model.}
    \label{tab:S4}
    \vspace{0.1in} \small
    \begin{tabular}{llcccccccccc}
      \hline
  & Dimension & &Sobel &JS & JT\_Comp & DACT &Bootstrap & AJS\\
      \hline
     $n=200$
   & $d=10$ &FWER &0.0030 &0.0108  & 0.1348 & 0       & 0.0124 & 0.0242\\
      &     &Power&0.0687 &0.2009  & 0.2116 & 0.0204  & 0.1713 & 0.2536\\
   & $d=15$ &FWER &0.0028 &0.0100  & 0.1480 & 0.0002  & 0.0104 & 0.0334\\
      &     &Power&0.0470 &0.1703  & 0.2049 & 0.0545  & 0.1297 & 0.2128\\
   & $d=20$ &FWER &0.0030 &0.0070  & 0.1564 & 0.0036  & 0.0076 & 0.0404\\
      &     &Power&0.0349 &0.1513  & 0.2118 & 0.1037  & 0.1003 & 0.1872\\
      \hline
     $n=500$
   & $d=10$ &FWER &0.0090 &0.0120  & 0.1550 & 0       & 0.0148 & 0.0286\\
      &     &Power&0.4104 &0.5449  & 0.4225 & 0.0128  & 0.5412 & 0.6546\\
   & $d=15$ &FWER &0.0048 &0.0082  & 0.1566 & 0       & 0.0106 & 0.0370\\
      &     &Power&0.3800 &0.5089  & 0.4175 & 0.0876  & 0.5030 & 0.6148\\
   & $d=20$ &FWER &0.0038 &0.0062  & 0.1776 & 0.0004  & 0.0092 & 0.0402\\
      &     &Power&0.3614 &0.4918  & 0.4306 & 0.2160  & 0.4851 & 0.5978\\
      \hline
     $n=1000$
   & $d=10$ &FWER &0.0098 &0.0112  & 0.1666 & 0       & 0.0146 & 0.0254\\
      &     &Power&0.6741 &0.7966  & 0.5256 & 0.0066  & 0.8020 & 0.8470\\
   & $d=15$ &FWER &0.0060 &0.0072  & 0.1606 & 0       & 0.0100 & 0.0370\\
      &     &Power&0.6330 &0.7689  & 0.5265 & 0.1056  & 0.7769 & 0.8274\\
   & $d=20$ &FWER &0.0028 &0.0034  & 0.1848 & 0       & 0.0078 & 0.0388\\
      &     &Power&0.6018 & 0.7490 & 0.5414 & 0.3100  & 0.7589 & 0.8146\\
      \hline
    \end{tabular}
  \end{center}
\end{table}

\clearpage
\begin{table}[htp] 
  \begin{center}
    \caption{The coverage probability and length of 95\% confidence interval with logistic mediation model$^\ddag$.}
    \label{tab:S5}
    \vspace{0.1in} \small
    \begin{tabular}{llcccccccccc}
      \hline
   &&&\multicolumn{3}{c}{CP} &&   \multicolumn{3}{c}{LCI} \\
      \cline{4-6}\cline{8-10}
      &$(\alpha_k, \beta_k)$ & &Sobel & Bootstrap & ASobel & & Sobel & Bootstrap&ASobel \\
      \hline
   $n=200$&
   (0, 0)      & &1      & 0.9976 & 0.9462 &&0.09662 & 0.18394 & 0.05037   \\
&  (0.35, 0)   & &0.9644 & 0.9396 & 0.9596 &&0.38635 & 0.60982 & 0.38482   \\
& (0.5, 0)     & &0.9508 & 0.9394 & 0.9508 &&0.54536 & 0.85866 & 0.54536   \\
&(0, 0.75)     & &0.9920 & 0.9614 & 0.9030 &&0.25726 & 0.38664 & 0.22188   \\
&(0, 0.8)      & &0.9870 & 0.9530 & 0.9072 &&0.27034 & 0.40778 & 0.24024   \\
&(0.35, 0.75)  & &0.9480 & 0.9182 & 0.9470 &&0.48502 & 0.94437 & 0.48440   \\
&(0.45, 0.85)  & &0.9526 & 0.9000 & 0.9526 &&0.60453 & 1.22664 & 0.60453  \\
      \hline
   $n=500$&
   (0, 0)      & &0.9998 & 0.9976 & 0.9514 &&0.03471 & 0.05171 & 0.01737   \\
&  (0.35, 0)   & &0.9526 & 0.9416 & 0.9526 &&0.22141 & 0.24178 & 0.22141   \\
& (0.5, 0)     & &0.9490 & 0.9450 & 0.9490 &&0.31457 & 0.34101 & 0.31457   \\
&(0, 0.75)     & &0.9710 & 0.9482 & 0.9334 &&0.14251 & 0.15533 & 0.13613   \\
&(0, 0.8)      & &0.9712 & 0.9502 & 0.9474 &&0.15098 & 0.16404 & 0.14711   \\
&(0.35, 0.75)  & &0.9492 & 0.9388 & 0.9492 &&0.27651 & 0.29945 & 0.27651    \\
&(0.45, 0.85)  & & 0.9482 &0.9328 & 0.9482 &&0.34346 & 0.37253 & 0.34346   \\
      \hline
   $n=1000$&
   (0, 0)      & &1      & 0.9994 &  0.9478 &&0.01686 & 0.02423 & 0.00843   \\
&  (0.35, 0)   & & 0.9528 & 0.9454 & 0.9528 &&0.15165 & 0.15741 & 0.15165   \\
& (0.5, 0)     & &0.9516 & 0.9496 & 0.9516 &&0.21613 & 0.22400 & 0.21613   \\
&(0, 0.75)     & &0.9616 & 0.9472 & 0.9564 &&0.09632 & 0.10023 & 0.09582   \\
&(0, 0.8)      & & 0.9598 & 0.9454 & 0.9582 &&0.10266 & 0.10656 & 0.10247  \\
&(0.35, 0.75)  & &0.9422 & 0.9392 & 0.9422 &&0.18969 & 0.19623 & 0.18969  \\
&(0.45, 0.85)  & &0.9498 & 0.9414 &0.9498  &&0.23585 & 0.24425 & 0.23585   \\
\hline
    \end{tabular}
  \end{center}
  {\vspace{0cm} \hspace{-0.3cm}\footnotesize $\ddag$ ``CP" denotes the empirical coverage probability; ``LCI" denotes the length of 95\% confidence interval.}
\end{table}

\clearpage\begin{table}[htp] 
  \begin{center}
    \caption{The coverage probability and length of 95\% confidence interval with Cox mediation model$^\ddag$.}
    \label{tab:S6}
    \vspace{0.1in} \small
    \begin{tabular}{llcccccccccc}
      \hline
   &&&\multicolumn{3}{c}{CP} &&   \multicolumn{3}{c}{LCI} \\
      \cline{4-6}\cline{8-10}
      &$(\alpha_k, \beta_k)$ & &Sobel & Bootstrap & ASobel & & Sobel & Bootstrap&ASobel \\
      \hline
   $n=200$&
   (0, 0)        & &0.9998 & 0.9984 & 0.9530 &&0.03376 & 0.05178 & 0.01757   \\
&  (0.35, 0)     & &0.9624 & 0.9494 & 0.9568 &&0.13767 & 0.15948 &  0.13704   \\
& (0.5, 0)       & &0.9488 & 0.9446 & 0.9488 &&0.19484 & 0.22181 & 0.19484   \\
&(0, 0.5)        & &0.9688 & 0.9484 & 0.9652 &&0.14885 & 0.16165 & 0.14850   \\
&(0, 0.45)       & &0.9746 & 0.9482 & 0.9644 &&0.13393 & 0.14657 &  0.13277   \\
&(0.45,  0.55)   & &0.9402 & 0.9464 & 0.9402 &&0.24669 & 0.27106 & 0.24669   \\
&(0.35, 0.35)    & &0.9390 & 0.9500 & 0.9390 &&0.17099 & 0.18929 &  0.17093   \\
      \hline
   $n=500$&
   (0, 0)        & &0.9998 & 0.9982 & 0.9490 &&0.01259 & 0.01822 & 0.00632   \\
&  (0.35, 0)     & &0.9538 & 0.9436 & 0.9538 &&0.08157 & 0.08544 & 0.08157   \\
& (0.5, 0)       & &0.9452 & 0.9382 & 0.9452 &&0.11612 & 0.12078 & 0.11612   \\
&(0, 0.5)        & &0.9554 & 0.9410 & 0.9554 &&0.09002 & 0.09254 & 0.09002   \\
&(0, 0.45)       & &0.9602 & 0.9458 & 0.9602 &&0.08146 & 0.08394 & 0.08146   \\
&(0.45,  0.55)   & &0.9506 & 0.9458 & 0.9506 &&0.14883 & 0.15322 & 0.14883   \\
&(0.35, 0.35)    & &0.9490 & 0.9498 & 0.9490 &&0.10299 & 0.10620 & 0.10299   \\
      \hline
   $n=1000$&
   (0, 0)        & &0.9996 & 0.9990 & 0.9504 &&0.00622 &0.00882 & 0.00311  \\
&  (0.35, 0)     & &0.9552 & 0.9492 & 0.9552 &&0.05656 & 0.05752 &0.05656   \\
& (0.5, 0)       & &0.9476 & 0.9424 & 0.9476 &&0.08074 & 0.08179 & 0.08074   \\
&(0, 0.5)        & &0.9518 & 0.9462 & 0.9518 &&0.06271 & 0.06350 & 0.06271  \\
&(0, 0.45)       & &0.9568 & 0.9500 & 0.9568 &&0.05659 & 0.05731 &  0.05659   \\
&(0.45,  0.55)   & &0.9490 & 0.9448 & 0.9490 &&0.10378 & 0.10486 &  0.10378   \\
&(0.35, 0.35)    & &0.9506 & 0.9494 & 0.9506 &&0.07145 &0.07224 & 0.07145  \\
\hline
    \end{tabular}
  \end{center}
  {\vspace{0cm} \hspace{-0.3cm}\footnotesize $\ddag$ ``CP" denotes the empirical coverage probability; ``LCI" denotes the length of 95\% confidence interval.}
\end{table}

\clearpage
\section{The manual for R package ``AdjMed"}
The following section presents detailed instructions on the practical implementation of the R package {\tt AdjMed}, including two functions {\tt AJS()} and {\tt ASobel()}. First we can install the {\tt AdjMed} from the GitHub with the following R codes:

{\tt > library(devtools)}

{\tt > devtools::install\_github("zhxmath/AdjMed")}\\
The R function {\tt AJS()} is used to perform adjusted joint significance test for mediation effect. The arguments when implementing the {\tt AJS()} are given as follows:

{\tt > AJS(X, M, Y, Z, Delta, Model)}\\

\begin{table}[htp] 
\caption{\label{Tab-S7} Overview of the arguments in functions {\tt AJS()} and {\tt ASobel()}}
\vspace{0.2cm}\centering
\begin{tabular}{lcp{11.5cm}}
\hline
Arguments &&  Description \\ \hline
{{\tt X}}   &&  a vector of exposures.  \\
{{\tt M}}   &&  a matrix of continuous mediators.  Rows represent samples, columns represent variables.\\
{{\tt Y}}   &&  a vector of observed outcomes. \\
{{\tt Z}}   &&   a matrix of covariates.  Rows represent samples, columns represent variables, {\tt Z= "null"} when the covariates are not available.  \\
{{\tt Model}}   &&  the type of outcome. {\tt Model= "Linear"} for continuous outcome; {\tt Model= "Logistic"} for binary outcome; {\tt Model= "Cox"} for time-to-event outcome with Cox model.  \\
{{\tt  Delta}}   &&  a vector of indicators for {\tt Model= "Cox"}, where 1=uncensored, 0=censored; {\tt Delta="null"} when {\tt Model= "Linear"} and {\tt Model = "Logistic"}.  \\
{{\tt tau}}   &&   the (1-tau)\% confidence level;  e.g., tau=0.05 denotes 95\% confidence level. The term is exclusively intended for the implementation of {\tt ASobel()}. \\
\hline
\end{tabular}
\end{table}

The {\tt ASobel()} is used to perform adjusted Sobel-type confidence interval for mediation effect, where the arguments are

{\tt > ASobel (X, M, Y, Z, Delta, Model,tau)} \\
In Tables \ref{Tab-S7} and \ref{Tab-S8}, we present the arguments and outputs of R functions {\tt AJS()} and {\tt ASobel()}.

\begin{table}[htp] 
\caption{\label{Tab-S8} The outputs of R functions {\tt AJS()} and {\tt ASobel()}}
\vspace{0.2cm}\centering
\begin{tabular}{lcp{11.5cm}}
\hline
Arguments &&  Description \\ \hline
{{\tt alpha\_est}}   &&  coefficient estimate of exposure (X) $\rightarrow$ mediator (M).  \\
{{\tt alpha\_SE}}   &&  the standard error for {\tt alpha\_est}. \\
{{\tt beta\_est}}   &&  coefficient estimate of mediator (M) $\rightarrow$  outcome (Y).  \\
{{\tt beta\_SE}}   &&   the standard error for {\tt beta\_est}. \\
{{\tt P\_AJS}}   &&   the p-values of mediation tests towards {\tt AJS()}. \\
{{\tt CI\_Asobel}}   &&  the (1-tau)\%  confidence intervals for mediation effects. \\
\hline
\end{tabular}
\end{table}
An illustrative R example of linear mediation model is provided as follows:
{\footnotesize
\begin{lstlisting}
library(MASS)
library(survival)
library(AdjMed)
p <- 5 # the dimension of mediators
q <- 2
n <- 500
alpha <- matrix(0,1,p) # the coefficients for X -> M
beta <- matrix(0,1,p) # the coefficients for M -> Y
alpha[1:3] <- 0.5
beta[1:3] <- 0.5
sigma_e <- matrix(0,p,p)
rou <- 0.25  # the correlation of  M
for (i in 1:p) {
  for (j in 1:p) {
    sigma_e[i,j]=(rou^(abs(i-j)));
  }
}


X <- matrix(rnorm(n, mean = 0, sd = 1),n,1) # expoure
zeta <- matrix(0.3,p,q) # the coefficients of covariates for X -> M
eta <- matrix(0.5,1,q) # the coefficients of covariates for M -> Y
gamma <- 0.5 # the direct effect
gamma_total <- gamma + alpha%*%t(beta) # the total effect
E <- matrix(rnorm(n, mean = 0, sd = 1),n,1)
mu <- matrix(0,p,1)
e <- mvrnorm(n, mu, sigma_e)
M <- 0.5+ X%*%(alpha) + e # the mediators
Y <-  0.5 + X*gamma + M%*%t(beta) + E # the response Y
\end{lstlisting}
{\tt fit\_AJS <- AJS(X, M, Y, Z="null", Delta="null", Model="Linear")\\
fit\_ASobel <- ASobel(X, M, Y, Z="null", Delta="null", Model="Linear",tau=0.05)\\
fit\_AJS\\
fit\_ASobel\\
}
}

\end{document}